\documentclass[12pt,preprint]{aastex}

\newcommand\oBV{\omega_{\rm BV}}
\newcommand\BVf{Brunt-V\"as\"ail\"a}
\newcommand\vecxi{\vec{\xi}}
\newcommand\tcool{t_{\rm cool}}
\newcommand\tgrow{t_{\rm grow}}

\newcommand\tcond{t_{\rm cond}}
\newcommand\rhodm{\rho_{\rm DM}}
\newcommand\calL{\mathcal{L}}
\newcommand\ksp{\kappa_{\rm Sp}}
\newcommand\DT{\Delta T}
\newcommand\DLr{\Delta L_r}
\newcommand\Msun{{\rm\,M_\odot}}

\newcommand\simgt{\lower.5ex\hbox{$\; \buildrel > \over \sim \;$}}
\newcommand\simlt{\lower.5ex\hbox{$\; \buildrel < \over \sim \;$}}

\slugcomment{Accepted for publication in ApJ}

\shortauthors{Kim \& Narayan}
\shorttitle{Thermal Instability in Galaxy Clusters}

\begin{document}

\pagebreak
\title{Thermal Instability in Clusters of Galaxies with Conduction}

\author{Woong-Tae Kim and Ramesh Narayan}
\affil{Harvard-Smithsonian Center for Astrophysics, \\ 
60 Garden Street, Cambridge, MA 02138}

\email{wkim@cfa.harvard.edu, narayan@cfa.harvard.edu}

\begin{abstract}
We consider a model of galaxy clusters in which the hot gas is in 
hydrostatic equilibrium and maintains energy balance between radiative
cooling and heating by thermal conduction.  We analyze the thermal
stability of the gas using a Lagrangian perturbation analysis.
For thermal conductivity at the level of $\sim20-40$\% of Spitzer 
conductivity, consistent with previous estimates for cluster gas, 
we find that the growth rate of the most unstable global radial mode 
is $\sim6-9$ times lower than the growth rate of local isobaric modes 
at the cluster center in the absence of conduction. The growth time in 
typical clusters is $\sim2-5$ Gyr, which is comparable to the time since 
the last major merger episode, when the gas was presumably well mixed.  
Thus, we suggest that thermal instability is not dynamically significant 
in clusters, provided there is an adequate level of thermal conduction.
On the other hand, if the heating of the gas is not the result of
thermal conduction or any other diffusive process such as turbulent
mixing, then the thermal instability has a growth time under a Gyr in
the central regions of the cluster and is a serious threat to
equilibrium.  We also analyze local nonradial modes and show that the
Lagrangian technique leads to the same dispersion relation as the
Eulerian approach, provided that clusters are initially in strict
thermal equilibrium.  Because cluster gas is convectively stable,
nonradial modes always have a smaller growth rate than equivalent
radial modes.
\end{abstract}


\section{Introduction}

Galaxy clusters contain a large amount of hot diffuse gas radiating
prolifically in thermal X-rays. In the absence of heat sources, the
radiative cooling due to this emission should induce a subsonic inflow 
of gas in the central regions, leading to substantial condensation of 
cold gas in rich clusters (e.g., \citealt{fab94} and references therein).  
However, recent high-resolution X-ray data from {\it XMM-Newton} and 
{\it Chandra} reveal no evidence for such cooling flows, nor any
significant evidence for mass dropout. In particular, there is no 
evidence for gas at temperatures below about a third of the average 
temperature \citep{pet01,pet03,tam01,boh01,fab01a,mol01,mat02,joh02}, 
suggesting that there must be some heat source (or sources) that 
prevents the gas from cooling below this limit.  Candidate heating 
mechanisms include radiative and mechanical power from active galactic 
nuclei (AGN) \citep{cio01,chu02,bru02,rey02,kai03}, thermal conduction 
from the hotter outer regions of the cluster to the center \citep[hereafter
ZN03]{nar01,fab02,gru02,zak03}, or perhaps both (e.g., ZN03,
\citealt{rus02,bri02,bri03}).

The effects of conduction in clusters have been widely discussed 
by many authors (e.g., \citealt{bin81, tuc83, ber86, bre88, gae89, ros89, 
dav92, pis96}; see \citealt{fab94} and references therein). 
It was shown that thermal conduction, when unimpeded, can
significantly reduce inferred mass deposition rates in cooling flow 
clusters \citep{ber86,ros89}, although such systems may end up being 
almost isothermal \citep{bre88}.
Nevertheless, thermal conduction was considered to be unimportant in 
galaxy clusters as it was believed that, in the presence of magnetic fields, 
the cross-field diffusion coefficient would be negligibly small.  
Since magnetic fields are ubiquitous in
clusters (e.g., \citealt{car02}), it appeared that the effective
isotropic conduction coefficient $\kappa$ in the presence of magnetic
fields would be smaller by orders of magnitude than the classical
\citet{spi62} value, $\ksp$, of an unmagnetized plasma.

\citet{nar01} recently showed that cross-field diffusion of electrons
is quite efficient if magnetic fields are fully turbulent and have a
wide range of coherence length scales.  Extending earlier work by
\citet{rec78}, \citet{cha98}, \citet{cha99}, and \citet{mal01}, and
adopting the model of \citet{gol95,gol97} for MHD turbulence, \citet{nar01}
estimated that $f\equiv\kappa/\ksp\sim0.2$ in a turbulent 
magnetized plasma.
Very recently, \citet{cho03} have used direct numerical simulations to
show that the turbulent diffusion of a scalar field in MHD turbulence
is as efficient as, or perhaps even more efficient than, the
prediction of \citet{nar01}.  These studies have led to the revival of
the idea that conduction might be an important heating source in
clusters.

The viability of thermal conduction as a heating mechanism
has found support in the work of \citet{voi02}, \citet{fab02}, and
ZN03, who showed that the level of conductivity needed to fit the
observed temperature distributions of X-ray clusters is consistent
with the theoretical estimate of \citet{nar01}.  In particular, ZN03
explicitly solved the equations for hydrostatic equilibrium and energy
balance between radiative cooling and conductive heating and showed
that conduction with $f\sim0.2-0.4$ provides reasonable profiles of
density and temperature for several clusters.  A similar result was
also obtained by \citet{bri03} using numerical
simulations.  Although other heating sources, including AGN jets,
cannot be ruled out, these studies suggest that an energy balance based
purely on conduction is not unreasonable.

While the above studies show that it is possible to construct
equilibrium cluster models with conductive and/or AGN heating, 
there is no guarantee that the equilibrium will be stable.  This
is because hot, optically-thin, X-ray-emitting gas is well-known to be
thermally unstable \citep{fie65}.  Since the growth time of the
thermal instability is comparable to the cooling time (e.g.,
\citealt{fab94}), one might expect rapid mass dropout as a result of
the instability, even when there is a source of heat to eliminate the
classic cooling flow.  On the other hand, the absence of any evidence
for gas below a few keV in clusters implies that the instability is
either absent or very slow.  A likely reason for the lack of
instability is thermal conduction, which is known to suppress thermal
instability on small scales (see, e.g.,
\citealt{fie65,def70,mal87,mck90}; ZN03).  However, all previous
analyses of this process have been limited to local WKB-type
perturbations, where the wavelength of the perturbations is much
smaller than the local radius of the system.  ZN03 found from their
WKB analysis the intriguing result that perturbations with wavelengths
up to almost the radius are stable, but that longer wavelengths are
probably unstable.  Such large-scale variations can be analyzed only
through a full global mode analysis, which has not been done so far.

In this paper we present a detailed and formal analysis of local and
global modes of thermal instability in equilibrium galaxy clusters
with conduction.  In \S2, we describe the basic equations we solve and
present equilibrium solutions for our fiducial model cluster. In \S3,
we carry out a linear stability analysis of the assumed equilibrium and
calculate the global radial modes of the system.  We show that there
is a single unstable mode, whose growth rate is much lower than the
usual thermal instability growth rate in the absence of conduction.
In \S4, we confirm the main results by means of numerical simulations
which show the development of the thermal instability under various
assumptions.  In \S5, we consider the stability of nonradial modes and
clarify a few points on which there has been confusion in the
literature.  We conclude in \S6 with a brief summary of the results
and a discussion of the implications.

\section{Perturbed Equations}

\subsection{Basic Equations}

We consider the thermodynamic evolution of a hot plasma subject to 
radiative cooling and thermal conduction. We do not consider any 
dynamical effects of magnetic fields. We also neglect the self-gravity 
of the gas which is generally weaker than the gravity of the dark 
matter.\footnote{We found that the explicit inclusion of self-gravity 
increases the growth rates of global thermal instability by only
$\sim3$ \% from the non-self-gravitating values.}
The governing hydrodynamic equations are  
\begin{equation}\label{eq_cont}
\frac{d\rho}{dt} + \rho\nabla\cdot\mathbf{v} = 0,
\end{equation}
\begin{equation}\label{eq_mome}
\frac{d\mathbf{v}}{dt} +\frac{1}{\rho}\nabla P + \nabla \Phi = 0,
\end{equation}
\begin{equation}\label{eq_engy}
\frac{1}{\gamma-1}\frac{dP}{dt} - 
\frac{\gamma}{\gamma-1} \frac{P}{\rho} \frac{d\rho}{dt} + 
\rho\calL + \nabla\cdot\mathbf{F} =0.
\end{equation}
Here, $d/dt \equiv \partial/\partial t + \mathbf{v}\cdot\nabla$
is the Lagrangian time derivative, 
$\rho$ is the mass density, $\mathbf{v}$ is the velocity, 
$T$ is the temperature, $\Phi$ is the gravitational potential,
$\gamma=5/3$ is the adiabatic index of the gas, 
$\rho\calL=2.1\times 10^{-27} n_e^2 T^{1/2}
~{\rm erg\;cm^{-3}\;s^{-1}}$ is the energy loss rate per unit volume
due to thermal bremsstrahlung (\citealt{ryb79}; ZN03),
and $P$ is the thermal pressure for which we adopt
the equation of state of an ideal gas,
\begin{equation}\label{eq_eos}
P = \frac{\rho k_B T}{\mu m_u} = \frac{\mu_e}{\mu} n_e k_B T,
\end{equation}
where $m_u$ is the atomic mass unit, $n_e$ is the
electron number density, and $\mu$ and $\mu_e$ denote 
the mean molecular weight per hydrogen atom and per electron,
respectively. We use $\mu=0.62$ and $\mu_e=1.18$, corresponding to
a fully ionized gas with hydrogen fraction $X=0.7$
and helium fraction $Y=0.28$ (ZN03).

In equation (\ref{eq_engy}), the conductive heat flux $\mathbf{F}$
is determined by 
\begin{equation}\label{eq_flux}
 \mathbf{F} = - \kappa \nabla T,
\end{equation}
where the conductivity $\kappa$ is a fraction $f$
of the classical \citet{spi62} conductivity $\ksp$,
\begin{equation}\label{eq_kappa}
\kappa = f\ksp = f \frac{1.84\times10^{-5}T^{5/2}}{\ln\Lambda_{\rm C}}
~{\rm erg\,s^{-1}\,K^{-1}\,cm^{-1}},
\end{equation}
with the Coulomb logarithm $\ln\Lambda_{\rm C}\sim 37$.
We do not consider nonlocal heat transport as in the model of \citet{chu93}.
In an unmagnetized plasma, conduction is isotropic and $f$ is unity.
The presence of a magnetic field generally reduces $f$ by 
resisting motions of thermal elections across the field lines.
Although the suppression of $\kappa$ would be very high if 
magnetic fields are uniform or only moderately tangled (e.g., \citealt{cha98}), 
\citet{nar01} showed that $f\sim0.2$ in a fully turbulent plasma medium 
in which magnetic fields are chaotic over a wide range of length scales.
In this paper, we assume that $f$ is constant in both space and time.

We use the NFW form of the dark matter distribution $\rhodm$  to
determine the gravitational potential $\Phi$ through
\begin{equation}\label{eq_poss}
\nabla^2 \Phi
= 4\pi G\rhodm(r) 
=\frac{2GM_0}{(r+r_c)(r+r_s)^2},
\end{equation}
with a softened core radius $r_c$, a scale radius $r_s$, and
a characteristic mass $M_0$. Utilizing the mass-temperature relation
of \citet{afs02} and the mass-scale relation of \citet{mao97},
we determine $M_0$ and $r_s$ for a given cluster from the observed 
temperature in the outer regions of the cluster (see ZN03).
For $r_c$, which determines the shape of the potential in the very inner 
parts, we adopt the best-fit values (either $r_c=0$ or $r_c=r_s$/20) 
recommended by ZN03.

Without any heating source to compensate for X-ray cooling, 
a cluster would lose its thermal energy at a rate $\rho\calL$.
Assuming that cooling occurs at fixed pressure,
we may define from equation (\ref{eq_engy}) the isobaric cooling 
time\footnote{Note that the cooling time as defined here is
$\mu_e/\mu$ times longer than the conventional definition, 
$5k_BT/2n_e\Lambda$, with $\Lambda=\rho\calL/n_e^2$ 
(cf.\ \citealt{sar88,dav01}).
At constant pressure, equation (\ref{eq_engy}) gives
$T(t)/T(0)=n_e(0)/n_e(t)=(1-3t/2\tcool)^{2/3}$.}
\begin{equation}\label{eq_tcool}
\tcool \equiv \frac{\gamma}{\gamma-1}\left(\frac{P}{\rho\calL}\right)
= 0.96\,{\rm Gyr}\,
\left(\frac{n_e      }{0.05\,{\rm cm^{-3}}}\right)^{-1}
\left(\frac{k_BT     }{2\,{\rm keV}}\right)^{1/2}.
\end{equation}
Similarly, we define the conduction time as
\begin{equation}\label{eq_tcond}
\tcond \equiv \frac{\gamma}{\gamma-1}
\left(\frac{\lambda^2 P}{4\pi^2\kappa T}\right),
\end{equation}
where $\lambda$ is the length scale over which the temperature changes 
appreciably. By comparing equations (\ref{eq_tcool}) and (\ref{eq_tcond}),
we see that conductive heating becomes comparable to radiative cooling at
$\lambda\sim\lambda_F$ where 
\begin{equation}\label{eq_flen}
\lambda_F \equiv 2\pi \left(\frac{\kappa T}{\rho\calL}\right)^{1/2}
= 31.4\,{\rm kpc}\,
\left(\frac{f}{0.2}\right)^{1/2}
\left(\frac{n_e      }{0.05\,{\rm cm^{-3}}}\right)^{-1}
\left(\frac{k_BT     }{2\,{\rm keV}}\right)^{3/2}
\end{equation}
is the Field length \citep{fie65,mck90}.
This corresponds to the length scale below which thermal conduction erases 
temperature perturbations completely \citep{fie65}. Since the length scale 
for temperature variations is typically $\sim10-100$ kpc, we conclude that 
thermal conduction with $f\simgt0.1$ can potentially support galaxy clusters 
against radiative cooling (see \citealt{nar01,fab02,gru02}; ZN03).

\subsection{Initial Equilibrium}

It is straightforward to find initial equilibrium solutions of 
equations (\ref{eq_cont})-(\ref{eq_poss}). We assume that the equilibrium 
is spherically symmetric, static, and time-independent. 
Working in spherical coordinates ($r$, $\theta$, $\phi$),
we simplify equations (\ref{eq_mome}), (\ref{eq_engy}), and 
(\ref{eq_flux}) to
\begin{equation}\label{eq_hse}
\frac{dP}{dr} = - \rho \frac{d\Phi}{dr},
\end{equation}
\begin{equation}\label{eq_ebal}
\frac{1}{r^2}\frac{d}{dr}\left(r^2F_r\right) = -\rho\calL,
\end{equation}
\begin{equation}\label{eq_fbal}
\kappa\frac{dT}{dr} = -F_r,
\end{equation}
where $F_r$ denotes the radial heat flux. 

Full numerical solutions of the differential equations (\ref{eq_poss}) and 
(\ref{eq_hse})-(\ref{eq_fbal}) were presented by ZN03. They varied $f$ 
to find the solutions that give the best fits to the observed density 
and temperature profiles of ten clusters. They found that thermal 
conduction with $f\sim 0.2-0.4$ explains the density and temperature 
distributions of five clusters fairly well (A1795, A1835, A2199, A2390, 
RXJ1347.5-1145). However, five other clusters require unphysically large 
values of $f$, indicating that those clusters are incompatible with a 
pure conduction model and require other heat sources. In this paper we 
take the best-fit parameters obtained by ZN03 for the five clusters that 
are consistent with conduction, and we analyze their stability to global 
radial modes. Table \ref{table1} lists the model parameters and various 
time scales. We adopt the cluster A1795 as our fiducial model.

\begin{figure}
\epsscale{1.}
\plotone{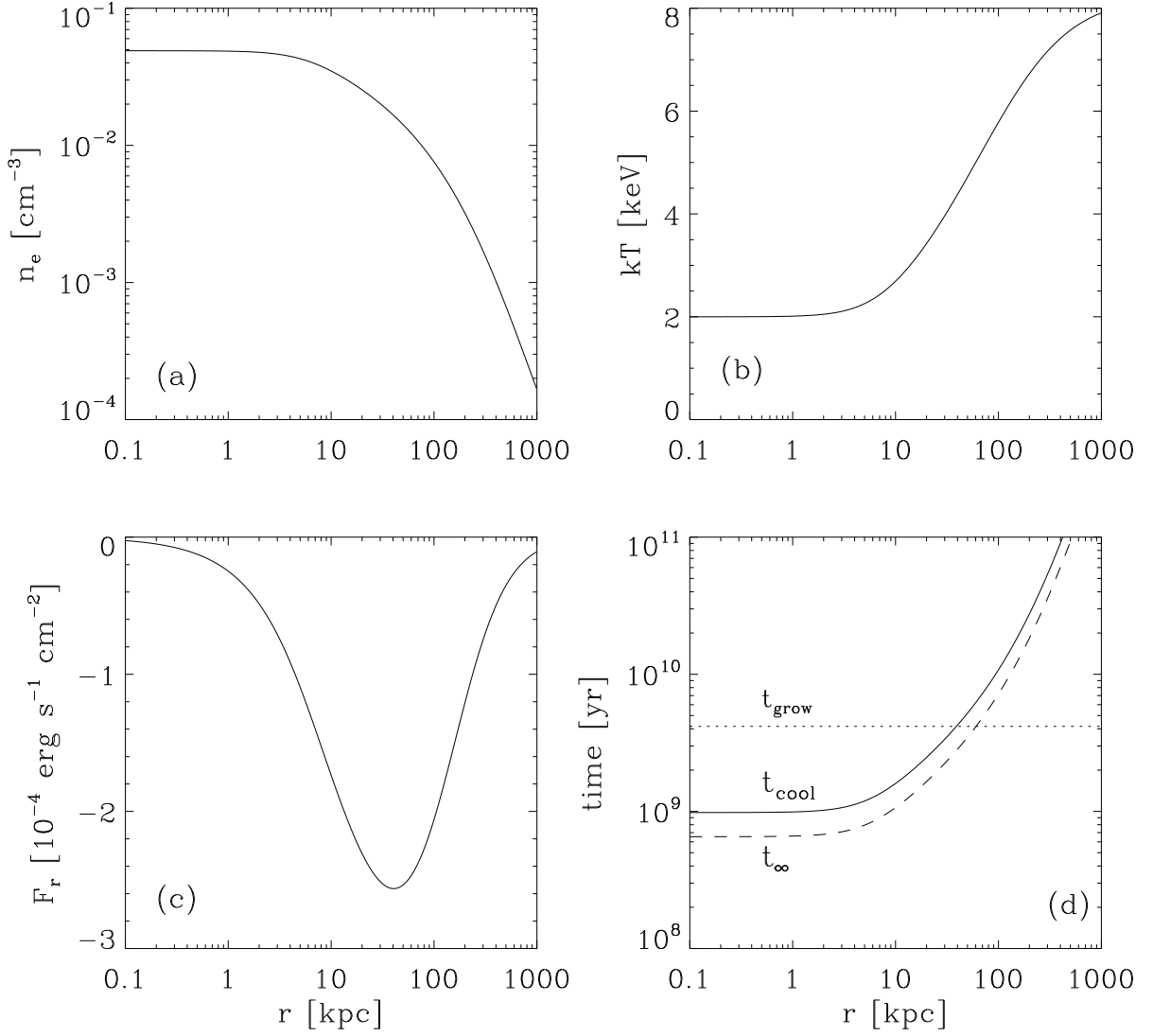}
\caption{($a$) Density, ($b$) temperature, ($c$) radial heat flux, and ($d$) 
isobaric cooling time ($\tcool$, solid line, see eq.\ [\ref{eq_tcool}]) 
of a model of A1795 with conduction.
The chosen parameters are $f=0.2$, $T(0)=2$ keV, 
$n_e(0)=0.049\,{\rm cm^{-3}}$, $M_0 = 6.6\times10^{14}\Msun$,
and $r_s=20r_c=460$ kpc (ZN03). Shown also in ($d$) are 
the growth times of local ($t_\infty$, dashed, see eq.\ [\ref{eq_tinf}]) and 
global (dotted) thermal instability.
\label{back_a1795}}
\end{figure}

Figure \ref{back_a1795} illustrates an equilibrium solution for A1795.
The temperature is minimum at $r=0$ and begins to rise sharply at 
$r\sim 5$ kpc. Expansion of the variables near the center gives 
$T(r)\sim T_0 + (\rho_0\calL_0/6\kappa_0)r^2$, where the subscript ``0'' 
indicates the values at the center. Therefore, stronger cooling or smaller 
conductivity would cause the temperature to increase faster. Notice that 
the temperature is always an increasing function of the radius. Radial heat 
influx is largest at $r\sim50$ kpc beyond which the low density makes 
cooling as well as conductive heating unimportant. 
Figure \ref{back_a1795}$d$ plots the local isobaric cooling time $\tcool$ 
defined in equation (\ref{eq_tcool}) (solid line) and the growth time, 
$t_\infty$ (see eq.\ [\ref{eq_tinf}]), of local isobaric thermal 
perturbations without conduction (dashed line), which are compared with 
the growth time of the global radial mode in the presence of conduction 
($\tgrow$; dotted line): 
we discuss these instability time scales in the next section. 

The radial equilibrium profiles of A1795 presented in Figure \ref{back_a1795}
are close to but not exactly the same as those derived by \citet{ett02}.
Since A1795 is known to contain a cool filament near the center
\citep{fab01b}, the real density and temperature distributions are 
somewhat non-axisymmetric. Our axisymmetric equilibrium model of A1795
should, therefore, be regarded as an idealized version of 
what is a much more complex distribution of the intracluster medium in A1795. 
As we will show below, however, the lowest-order radial mode of thermal 
instability is the fastest growing mode, so that neglecting the 
non-axisymmetric parts in the background profiles probably does not 
affect the results significantly.

\section{Linear Analyses of Radial Modes}
\subsection{Lagrangian Perturbations}

We linearize equations (\ref{eq_cont})-(\ref{eq_flux}) in the Lagrangian 
framework. A Lagrangian perturbation, represented by an operator $\Delta$, 
is related to an Eulerian perturbation $\delta$ in the usual way,
\begin{equation}\label{oper}
\Delta = \delta + \vecxi\cdot\nabla,
\end{equation}
where the vector $\vecxi$ measures the Lagrangian displacement of a fluid 
element from its unperturbed location (e.g., \citealt{sha83}). One of the 
advantages of adopting the Lagrangian approach is that it simplifies the 
perturbed equations greatly. For instance, the perturbed continuity 
equation (\ref{eq_cont}) becomes
\begin{equation}\label{eq_Lcont}
\Delta\rho = -\rho\nabla\cdot\vecxi.
\end{equation}
Nevertheless, both Lagrangian and Eulerian descriptions should
give the same results, especially if the initial state is static and 
in complete equilibrium.
Various properties of $\Delta$ and the commutation relations associated
with $\Delta$ and $\delta$ can be found in \citet{sha83}.

We assume that the perturbations are all radial.  Applying $\Delta$ to 
equations (\ref{eq_mome})-(\ref{eq_flux}) and using 
equation (\ref{eq_Lcont}), we obtain
\begin{equation}\label{eq_Lmome}
\frac{d^2\xi_r}{dt^2} = 
\frac{P}{\rho}\frac{\partial}{\partial r}\left(\frac{1}{r^2}
\frac{\partial r^2\xi_r}{\partial r}\right)
- \frac{1}{\rho}\frac{\partial }{\partial r}
\left(P\frac{\DT}{T}\right) - 
\frac{\partial}{\partial r}\left(\xi_r\frac{d\Phi}{dr}\right),
\end{equation}
\begin{equation}\label{eq_Lengy}
\frac{1}{4\pi r^2}\frac{\partial}{\partial r} \DLr
=\left( \frac{P}{\gamma-1}\frac{d}{dt} + \rho T \calL_T \right)
\frac{\DT}{T} +
\left(P\frac{d}{dt} - \rho^2\calL_\rho\right) (\nabla\cdot\vecxi),
\end{equation}
\begin{equation}\label{eq_flux_r}
\kappa T \frac{\partial}{\partial r}\left(\frac{\DT}{T}\right)
= F_r\left(\frac{\Delta\kappa}{\kappa}+\frac{\DT}{T} -\frac{d\xi_r}{dr}
+ 2\frac{\xi_r}{r}\right) + \frac{\DLr}{4\pi r^2},
\end{equation}
where $\calL_T\equiv \partial\calL/\partial T|_\rho$,
$\calL_\rho\equiv \partial\calL/\partial\rho|_T$, $\xi_r$ is the 
radial component of $\vecxi$, and $L_r\equiv - 4\pi r^2 F_r$
is the radial heat luminosity. 
We use $\xi_r$, $\DT$, and $\DLr$ as independent variables.

We seek linear eigenmodes that behave as $\sim e^{\sigma t}$ with time.
We then rewrite equations (\ref{eq_Lmome})-(\ref{eq_flux_r}) as
\begin{equation}\label{eq_f1}
\frac{d^2}{dr^2}\left(\frac{\xi_r}{r}\right) 
+ \left(\frac{4}{r} + \frac{d\ln P}{dr}\right)
\frac{d}{dr}\left(\frac{\xi_r}{r}\right)
+\frac{\rho}{P}\left(\frac{1}{r}\frac{d\Phi}{dr}
-4\pi G\rhodm - \sigma^2\right)\frac{\xi_r}{r} =
\frac{1}{rP} \frac{d}{dr}\left(P\frac{\DT}{T}\right), 
\end{equation}
\begin{equation}\label{eq_f2}
\frac{1}{4\pi r^2}\frac{d}{dr} \DLr
=\left( \frac{P\sigma}{\gamma-1} + \rho T \calL_T \right)
\frac{\DT}{T} +
\left(P\sigma - \rho^2\calL_\rho\right) 
\left[r\frac{d}{dr}\left(\frac{\xi_r}{r}\right) + 3\frac{\xi_r}{r}\right],
\end{equation}
\begin{equation}\label{eq_f3}
\kappa T \frac{d}{dr}\left(\frac{\DT}{T}\right)
= F_r\left[\frac{7}{2}\frac{\DT}{T} -r\frac{d}{dr}
\left(\frac{\xi_r}{r}\right)
+ \frac{\xi_r}{r}\right] + \frac{\DLr}{4\pi r^2},
\end{equation}
which are our desired perturbation equations. This set of ordinary 
differential equations forms an eigenvalue problem for global modes, 
with $\sigma$ as the eigenvalue, which can be solved numerically subject to 
appropriate boundary conditions. Before considering the full problem, 
we study local modes that do not depend on boundaries to gain some 
physical insight.

\subsection{Local Radial Modes}

Let us consider local WKB perturbations of the form 
$\sim e^{ik_rr+\sigma t}$ and let us assume that the radial wavenumber 
$k_r$ satisfies $k_rr\gg1$, $k_r(d\ln P/dr)\gg1$. We may then neglect 
the spherical geometry and ignore local gradients of the background 
quantities relative to spatial gradients of the perturbed variables.
We also assume that $\sigma^2 < Pk_r^2/\rho$, corresponding to slowly
evolving perturbations; this eliminates sound waves from consideration.
In this local approximation, equation (\ref{eq_f1}) simplifies to
\begin{equation}\label{eq_loc1}
ik_r\xi_r = \frac{\DT}{T},
\end{equation}
while  equations (\ref{eq_f2}) and (\ref{eq_f3}) may be combined to give
\begin{equation}\label{eq_loc2}
-(P\sigma - \rho^2\calL_\rho - \rho\calL)ik_r\xi_r
=\left(\frac{P\sigma}{\gamma-1}+\rho T\calL_T+k_r^2\kappa T\right)
\frac{\DT}{T}.
\end{equation}
Eliminating $\xi_r$ and $\DT/T$ from equations (\ref{eq_loc1}) and
(\ref{eq_loc2}), we find
\begin{equation}\label{ldisp}
\sigma = \sigma_\infty - \frac{\gamma-1}{\gamma}\frac{\kappa T}{P}k_r^2,
\end{equation}
where $\sigma_\infty$, defined by
\begin{equation}\label{eq_siginf}
\sigma_\infty \equiv \frac{\gamma-1}{\gamma P}
(\rho^2\calL_\rho+\rho\calL-\rho T\calL_T)
= - \frac{\gamma-1}{\gamma}
\frac{\rho T}{P}\left(\frac{\partial\calL/T}{\partial T}\right)_P,
\end{equation}
is the growth rate of isobaric thermal perturbations without conduction
\citep{fie65}.

Equation (\ref{ldisp}) is a local dispersion relation for thermal
fluctuations.\footnote{Equation (\ref{ldisp}) can be obtained by 
taking the limit of $k_r \gg \rho\sigma^2/P$ in 
the dispersion relation (15) of \citet{fie65} for a uniform medium. 
Note, however, that the condition $\calL=0$ in his initial 
equilibrium state gives 
$\sigma_\infty = -[(\gamma-1)\rho/\gamma P](\partial\calL/\partial T)_P$
instead of equation (\ref{eq_siginf}).}
In the absence of conduction, $\sigma=\sigma_\infty$, so that 
instability develops if the cooling function satisfies the 
generalized Field criterion
\begin{equation}\label{eq_cri}
\left(\frac{\partial\calL/T}{\partial T}\right)_P <0,
\end{equation}
for isobaric thermal instability \citep{bal86}. For X-ray emitting clusters
of galaxies with $\calL\propto\rho T^{1/2}$, the condition (\ref{eq_cri})
is easily met. The corresponding growth time amounts to
\begin{equation}\label{eq_tinf}
t_\infty \equiv \sigma_\infty^{-1} = \frac{2}{3}\tcool, 
\end{equation}
suggesting that local radial disturbances will grow
slightly faster than the isobaric cooling time of the system. 
Figure \ref{back_a1795}$d$ plots $t_\infty$ for A1795 as a dashed line.

It is well known that thermal conduction stabilizes short-wavelength 
perturbations against thermal instability
(e.g., \citealt{fie65,mal87,mck90}). It not only reduces growth
rates but also suppresses thermal instability completely if
$k_r > (3/2)^{1/2} k_F$, where $k_F=2\pi/\lambda_F$. 
The existence of unstable modes, therefore, requires large
wavelength perturbations.
ZN03 showed that the clusters they studied are thermally stable for all 
wavelengths up to approximately the radius. The fate of modes with
longer wavelengths was unclear in their study 
since a local analysis is no longer valid.
A global analysis is required to study such large scale perturbations.
This is the topic of the next subsection.

\subsection{Global Solutions}

To analyze the global stability problem,
we solve equations (\ref{eq_f1})-(\ref{eq_f3}) numerically.
Since  these equations are equivalent to
four first-order differential equations, we need to 
specify four boundary conditions.
The two inner boundary conditions are
\begin{equation}\label{innerb}
\frac{d\xi_r/r}{dr} = 0\;\;\;{\rm and}\;\;\;\frac{\DLr}{4\pi r^2} = 0,
\;\;\;{\rm at}\;r=0.
\end{equation}
The first condition guarantees that the solutions are regular, while 
the second condition corresponds to a zero gradient in the perturbed 
temperature at the center. Since $\xi_r=0$ at $r=0$ for radial modes, 
we fix $\xi_r/r=1$ as a normalization condition. 

The two other boundary conditions come from the outer boundary which 
we arbitrarily locate at $r_b=1$ Mpc. Since  all eigenfunctions that 
we have obtained are found to decay rapidly with increasing radius, 
the solutions are quite insensitive to the particular choice of the outer 
boundary conditions as well as the location of the outer boundary. 
This is because the cooling time well exceeds the Hubble time in the 
outer parts of clusters (Fig.\ \ref{back_a1795}$d$), and so the 
thermal instability develops very slowly near the 
outer boundary. The results presented in this paper correspond to
the specific boundary conditions
\begin{equation}\label{outerb}
\xi_r=0 \;\;\;{\rm and}\;\;\; \DT=0,
\;\;\;{\rm at}\;r=r_b.
\end{equation}
We have tried various other outer boundary conditions such as 
a fixed pressure, a fixed entropy, etc. The results are unchanged. 

The numerical calculations proceed as follows.
We first fix $\sigma$ and integrate equations (\ref{eq_f1})-(\ref{eq_f3}) 
from $r=0$ to $r=r_b$, setting $\alpha\equiv\DT/T|_{r=0}$ 
to an arbitrary value. At the outer boundary, we check the first condition 
in equation (\ref{outerb}), update $\alpha$ using the Newton-Raphson method,
and continue iterating until convergence is attained. 
Then, we scan $\sigma$ in the range 
$10^{-6}<|\sigma|r_b/c_0< 10^2$, where $c_0$
is the adiabatic sound speed at the cluster center,
and use the second condition in equation (\ref{outerb}) as a discriminant
for solutions. By this procedure, we do not miss any eigensolution.
We allow for both real and complex $\sigma$. 

\begin{figure}
\epsscale{1.}
\plotone{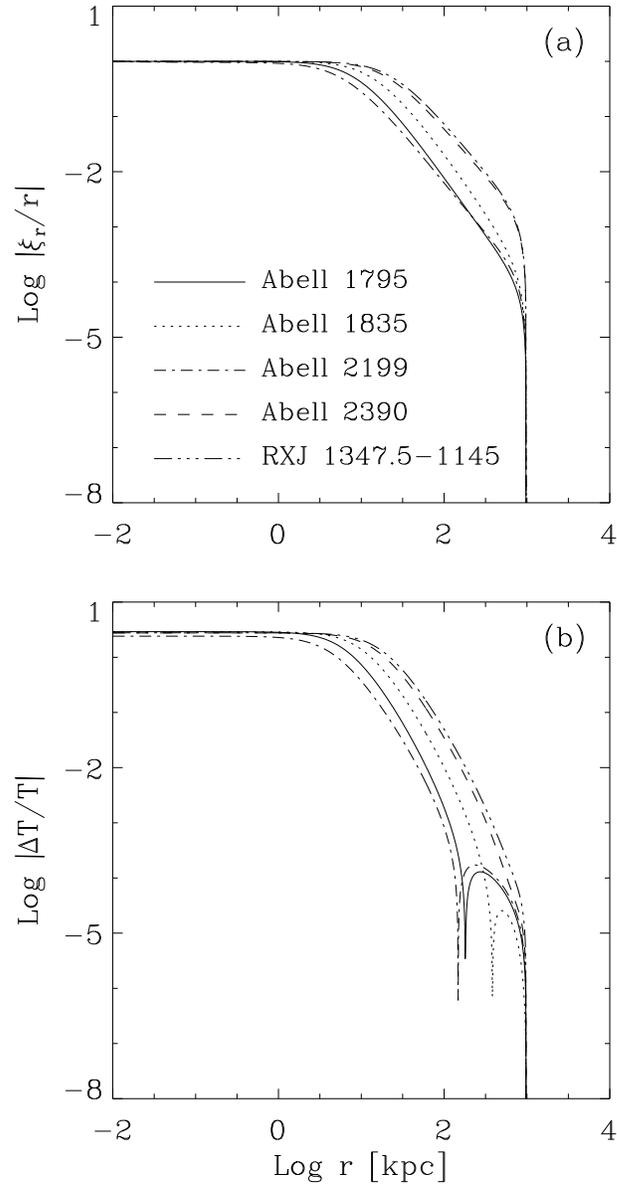}
\caption{Eigenfunctions of the radial unstable mode for 
five clusters, plotted as functions 
of radius. Note that the amplitudes of the solutions are 
largest near the center and decrease very rapidly as $r$ increases.
\label{eig_all}}
\end{figure}

When $\sigma$ is real and positive, we find that a cluster with 
$f\simgt0.2$ has only one unstable global mode. 
In Figure \ref{eig_all}, we plot the eigenfunctions of the unstable mode 
for all five clusters listed in Table \ref{table1}. The perturbations 
have largest amplitude near the center ($r<1$ kpc) and decay rapidly 
with increasing radius. This is consistent with the notion that the 
central regions have a much shorter cooling time and are thus more prone 
to thermal instability. Note that $\xi_r/r$ has the same sign as $\DT/T$, 
indicating that the thermal instability leads to a mass inflow.
We will show in \S4 that the mass inflow rate driven by the instability is, 
however, much lower than that in conventional cooling flows. 
As the last two columns of Table \ref{table1} shows, 
the growth time of the most unstable global radial modes is 
typically $\sim 2-5$ Gyr, 
which is about 6 to 9 times longer than that of 
local isobaric modes at the cluster center in the absence of conduction.
This time scale is not much shorter than the age of the system since the 
last major merger.  
Therefore, thermal instability in the presence of conduction is unlikely 
to be a serious threat to clusters.

Why is the growth time $\tgrow$ of the mode so much longer than the
growth time $t_{\infty,0}$ of the local isobaric mode at the cluster
center? The reason is that the eigenfunction extends over a considerable
range of $r$, of order tens of kpc (Fig.\ \ref{eig_all}). 
Over this range, $t_\infty$ increases from  $t_{\infty,0}$ at the
center to much larger values (Fig.\ \ref{back_a1795}$d$). Since
the growth time of the mode corresponds to an appropriate 
eigenfunction-weighted volume-average of $t_\infty$, and since the
averaging is dominated by $r\sim$ tens of kpc, this causes 
$\tgrow$ to be significantly larger than $t_{\infty,0}$.

\begin{figure}
\epsscale{1.}
\plotone{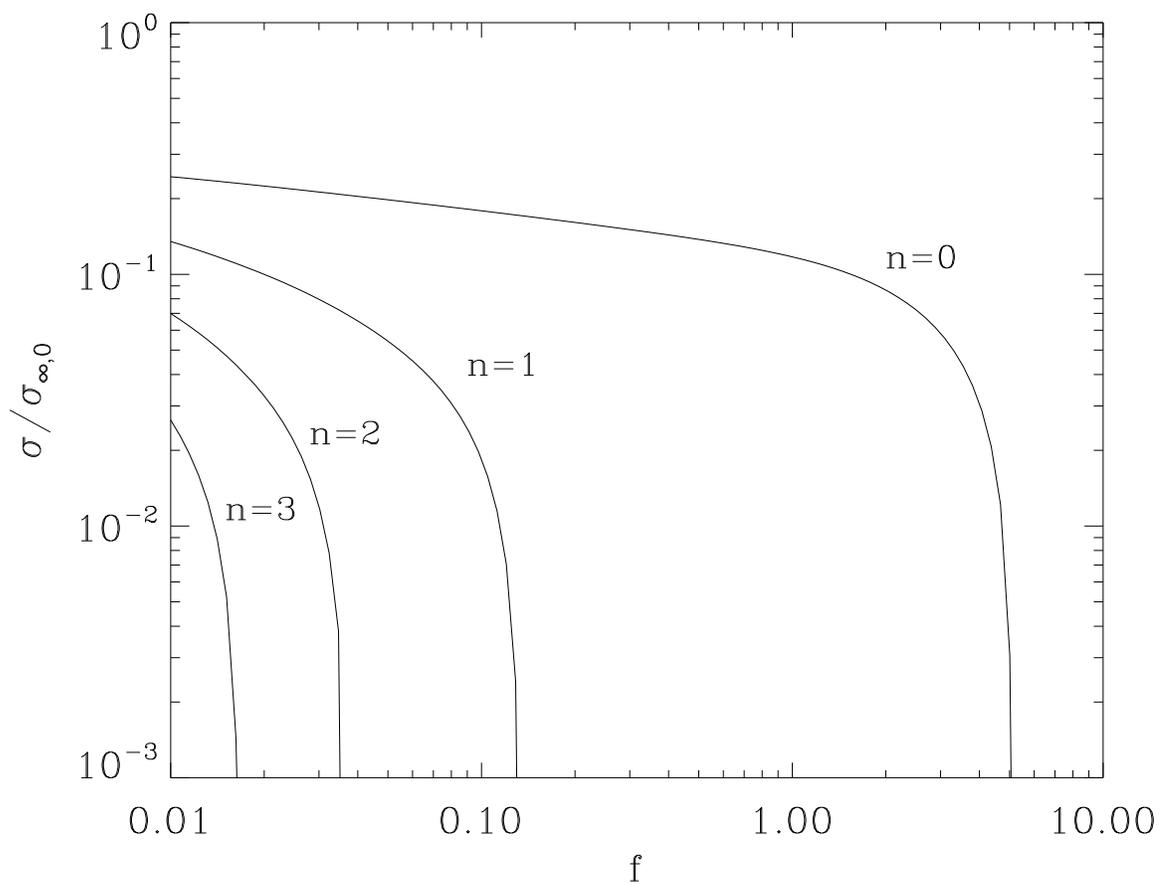}
\caption{Effect of thermal conductivity on the growth rates of
global unstable modes in A1795. The abscissa is the reduction 
factor $f$ of thermal conductivity applied to perturbations, 
while the ordinate is the eigenfrequency normalized by 
$\sigma_{\infty,0} = (0.64$ Gyr)$^{-1}$, the growth rate of the
local isobaric instability at the cluster center. Each curve is labeled by 
$n$,  the number of nodes in the corresponding eigenfunction $\xi_{r,n}/r$.  
See text for details.
\label{f_a1795_pos}}
\end{figure}

How effectively does conduction stabilize thermal instability?
To address this question, we have solved equations (\ref{eq_f1})-(\ref{eq_f3})
using different values of $f$ for the perturbations. We adopt the same 
density and temperature distributions as shown in Figure \ref{back_a1795}, 
which means that we take the same $f=0.2$ for the initial unperturbed state.
However, we allow $f$ to vary in the perturbations.  
In Figure \ref{f_a1795_pos}, we show the resulting mode growth rates $\sigma$
relative to $\sigma_{\infty,0}$, the central value of the growth rate of 
local thermal instability without conduction (see eq.\ [\ref{eq_siginf}]). 
When $f>5$ (although unphysical), we find that conduction quenches thermal
instability completely; that is, there is no unstable mode. For $0.14<f<5$, 
there exists a single unstable mode that has no node in the eigenfunction 
for $\xi_{r}/r$. The case shown in Figure \ref{eig_all}, $f=0.2$, corresponds 
to this range. 
When $f$ is below 0.14, a new mode appears and now the system 
has two unstable modes. Because the new mode has a node in $\xi_r/r$
somewhere inside the cluster, its growth rate is smaller than the zero-node 
mode, as equation (\ref{ldisp}) suggests. As $f$ keeps decreasing, new modes 
having lower growth rates successively emerge. In the limit of zero conduction,
there is an infinite number of unstable modes, with the maximum growth
rate equal to $\sigma_{\infty,0}$.

Figure \ref{f_a1795_pos} shows that over much of the range of $f$, 
the growth rate increases very slowly as $f$ decreases. As before,
the reason may be traced to the fact that the eigenfunction 
extends over a range of $r$ and $\sigma$ represents a suitable
average of $\sigma_\infty$ over this volume. Thus, even for $f=0.01$, 
the fundamental $n=0$ mode has a growth rate about 4 times smaller than 
$\sigma_{\infty,0}$. Eigenfunctions become more centrally concentrated as $f$ 
decreases, and thus a mode with smaller $f$ generally has a larger growth 
rate, though the increase of $\sigma$ with decreasing $f$ is rather slow 
(Fig.\ \ref{f_a1795_pos}).

\begin{figure}
\epsscale{1.}
\plotone{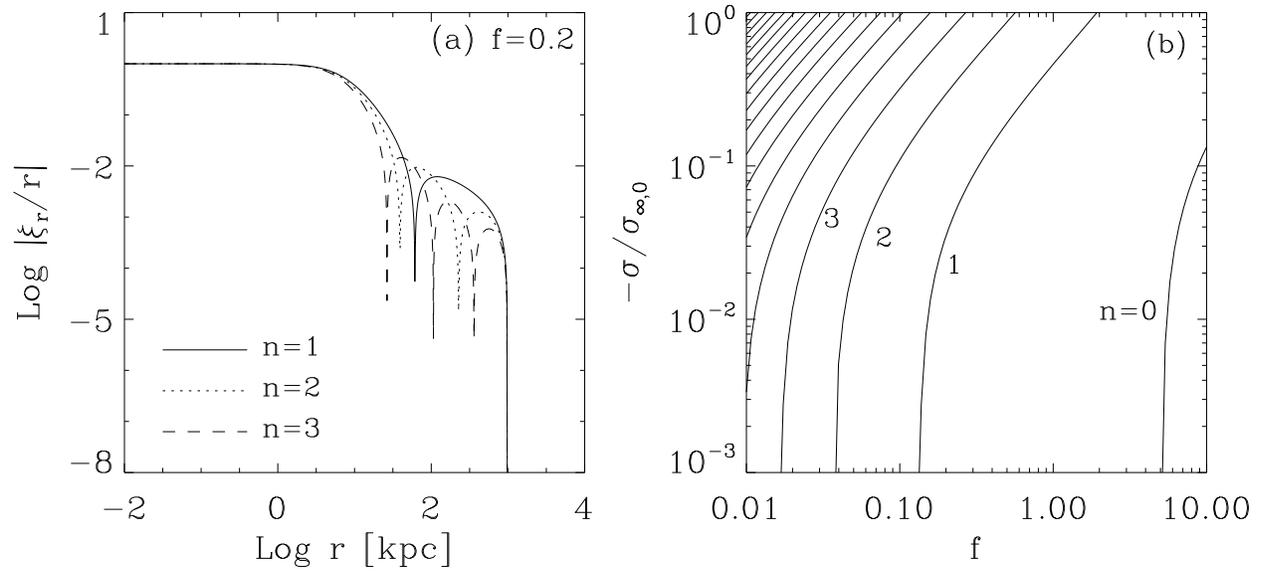}
\caption{ ($a$) Examples of eigenfunctions for decaying modes ($\sigma<0$)
in A1795 with $f=0.2$ ($b$) The effect of thermal conductivity 
on the decay rates of modes. Higher-order modes have larger decay rates.
\label{f_a1795_neg}}
\end{figure}

In addition to unstable modes for which thermal conduction plays a 
stabilizing role, clusters also contain decaying modes 
with real and negative $\sigma$. 
Figure \ref{f_a1795_neg} 
displays examples of eigenfunctions for a few selected decaying modes 
and shows the dependence of $\sigma$ on the reduction factor $f$.
In contrast to unstable modes, which have smaller frequencies as the number 
of nodes increases, higher-order decaying modes have larger (negative)
frequencies. This is consistent with equation 
(\ref{ldisp})\footnote{Since $\sigma\propto-\kappa Tk_r^2$ for decaying modes,
equation (\ref{ldisp}), which assumes $\sigma^2\ll Pk_r^2/\rho$,
does not describe higher-order modes very accurately.},
which implies that the presence of decaying modes is a simple 
manifestation of diffusion ironing out temperature perturbations.

We have searched for global solutions with complex $\sigma$, i.e., 
overstable modes, but found no overstable radial mode with Re($\sigma)>0$
under the imposed boundary conditions.
Physically, thermal overstability occurs when sound waves
are amplified by absorbing (losing) heat during the compression
(rarefaction) phase of oscillations \citep{fie65}.
Since the cooling function for X-ray emitting clusters
does not satisfy the isentropic instability criterion of \citet{fie65},
the absence of radial overstability is in fact guaranteed.
This leaves pure unstable modes as the sole growing modes.
Although some underdamping modes (complex $\sigma$ with negative real
parts) were found, they are of no interest and we do not discuss them
further.

\section{Nonlinear Evolution of Radial Modes}

As a check of the global linear stability analyses described in 
the previous section, we have solved the full 
dynamical equations (\ref{eq_cont})-(\ref{eq_poss}) using a 
time-dependent approach.
Starting with the equilibrium density and temperature profiles of A1795
calculated in \S2.2, we have run three models: 
(model A) a cluster without any heating;
(model B) a cluster with no conduction, but
constantly heated (by some fictitious agency) at a rate such as to 
maintain thermal balance in the initial equilibrium;
(model C) a cluster with self-consistent conductive heating with $f=0.2$.
Model A simulates the standard cluster cooling flow problem, whereas
models B and C are intended to explore the growth of thermal instability 
from an initial equilibrium state, 
with and without thermal conduction.
Note that our adopted cooling function for X-ray free-free emission
becomes invalid when the gas in models A-C cools below $\sim 1$ keV.
Nevertheless, these simulations allow us to 
confirm the growth rates of thermal instability that we computed
in \S3 and to estimate mass inflow rates resulting from either radiative 
cooling or thermal instability.

We follow the nonlinear evolution of our model clusters using the ZEUS 
hydrodynamic code \citep{sto92}. We construct a logarithmically-spaced radial 
grid, with 500 zones, from $r=1$ kpc to 1 Mpc, and carry out 
a one-dimensional simulation. We implement a fully implicit algorithm 
for thermal conduction in the energy equation \citep{pre92}.
All the variables at both the inner and outer boundaries are assigned to
have a vanishing gradient across the boundaries, except for 
the radial velocity. We fix the radial velocity to be zero
at the outer boundary, while allowing it to vary as a 
linear function of radius at the inner boundary.
On the initial equilibrium, we add as a perturbation the most 
unstable global eigenfunction for density that we found in \S3.3, 
with an amplitude equal to 0.1\% of the background density at the 
cluster center.

\begin{figure}
\vspace{-1cm}
\epsscale{1.}
\plotone{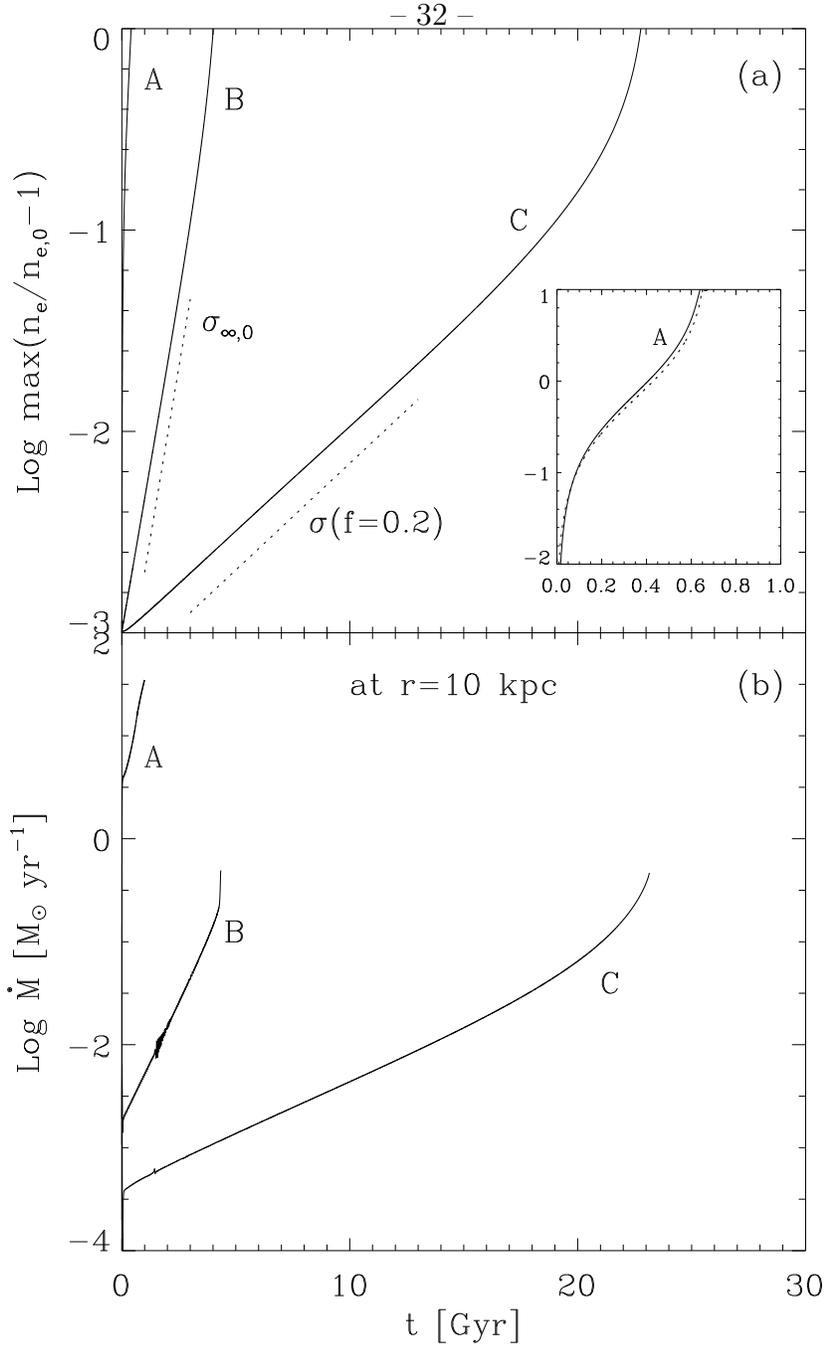}
\caption{Time evolution of ($a$) maximum density and ($b$) 
mass inflow rate measured at $r=10$ kpc in models A-C. 
The two dotted lines in ($a$) represent the maximum growth rates, 
$\sigma(f=0.2)$ = (4.1 Gyr)$^{-1}$
and $\sigma_{\infty,0}$ = (0.64 Gyr)$^{-1}$,
of thermal instability with and without conduction as estimated from
the linear analysis. As expected, these lines agree well with the
time evolutions of Models B and C.
The insert in ($a$) compares the evolution of 
the maximum density in model A (solid line) with the analytic prediction,
$n_e(t)/n_e(0)=(1-3t/2\tcool)^{-2/3}$, for isobaric cooling (dotted line).
See text for details.
\label{f_nevol}}
\end{figure}

\begin{figure}
\vspace{-1cm}
\epsscale{1.}
\plotone{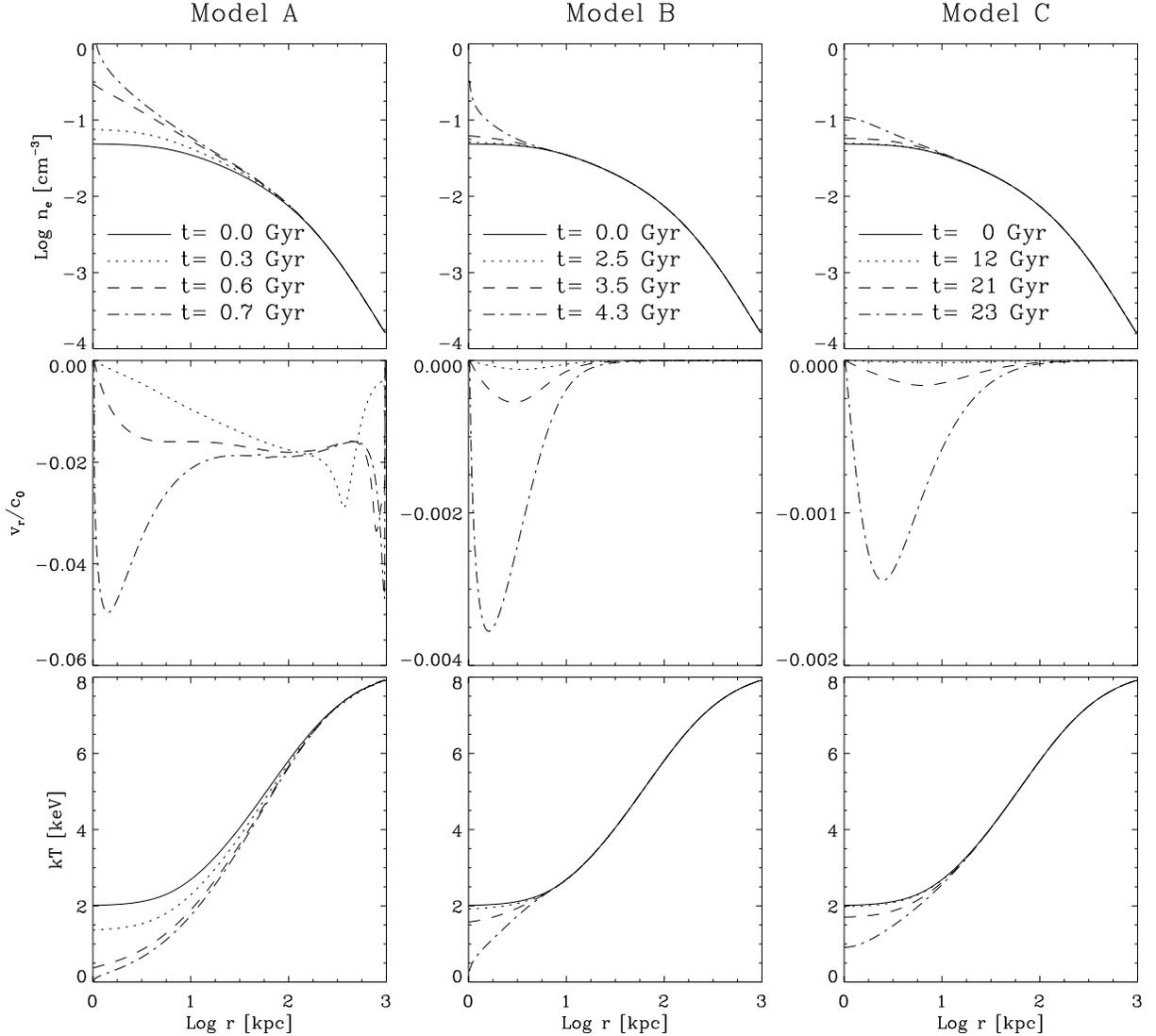}
\caption{Evolution of ({\it top row}) electron number density,
({\it middle row}) radial velocity in units of the central sound speed
$c_0=390$ km s$^{-1}$, and ({\it bottom row}) temperature, for models A-C.
{\it Left}: Strong radiative cooling in model A quickly leads to
an inward cooling flow, which causes the central density to increase by more
than three orders of magnitude in 0.9 Gyr.
{\it Middle}: Model B, which is initially in equilibrium but has no 
conduction, has perturbations that grow at a rate $\sigma_\infty$, 
driving the system into a highly nonlinear state in less than 5 Gyr.
{\it Right}: Thermal conduction in model C delays the development of 
thermal instability significantly. Even after 20 Gyr, the density
and temperature profiles are not very different from the initial ones.
\label{f_profiles}}
\end{figure}

Figure \ref{f_nevol} shows the evolutionary histories of the maximum 
density and the mass inflow rate $\dot{M}\equiv -4\pi r^2\rho v_r$ at 
$r=10$ kpc for models A-C. 
The two dotted lines in Figure \ref{f_nevol}$a$ represent the
analytic estimates for the growth rates of thermal instability 
with and without conduction; the predictions are in excellent agreement with 
the results of the numerical simulations.
We have also compared the perturbed density and velocity profiles 
in the linear regime of model C with the corresponding eigenfunctions 
obtained from the linear theory. Even though the radial grid does not 
cover the inner 1 kpc, they agree to within $\sim3$\%.
Radial profiles of electron number
density, radial velocity, and temperature for a few selected
epochs are plotted in Figure \ref{f_profiles}.

Model A, which is not in equilibrium and is subject to strong radiative 
cooling, immediately experiences radial inflow of mass everywhere.
Although the temperature of the cluster decreases steadily as the gas cools,
the thermal pressure does not drop owing to adiabatic compression,
which in turn maintains the radial velocity at the
level of about $\sim1-2$\% of the central sound speed until
$t\sim0.6$ Gyr. During this period, we find $n_e\propto T^{-1.05}$ so 
that the cooling flow is nearly isobaric. As the insert in Figure 
\ref{f_nevol}$a$ shows, the maximum density follows the prediction 
of isobaric cooling fairly well. 
When the cooling modifies the density and temperature 
distributions significantly 
($t>0.6$ Gyr), the increased cooling rate in the central parts
induces larger mass inflows,
causing the central density to increase in a runaway fashion. 
The change of density and temperature is so rapid that there
is no chance for thermal instability to grow.
We stop the simulation of model A at $t=0.85$ Gyr when the
central temperature becomes vanishingly small. The mass inflow rate
depends rather sensitively on time and is a linear function of 
radius for $r<250$ kpc, with a slope of $\sim30 \Msun$ yr$^{-1}$ kpc$^{-1}$,
at the end of the run (see Figure \ref{f_nevol}$b$).

Models B and C are initially in equilibrium with heating 
balancing cooling, but they both experience thermal instability. 
Without conduction, perturbations in model B grow at a rate of
$\sigma_{\infty,0} = (0.64$ Gyr)$^{-1}$ and become fully unstable
within $\sim4$ Gyr. 
Although the growth time of thermal instability in model B is slighter 
shorter than the isobaric cooling time in model A (see eq.\ [\ref{eq_tinf}]), 
model B takes a longer time for eventual runaway than model A.
This is simply because the former requires the growth of perturbations 
from a small amplitude, while the latter changes its background state 
immediately. 

Model C clearly illustrates the stabilizing effect of thermal conduction.
With $f=0.2$, conduction causes the maximum growth rate of thermal 
instability to be about six times smaller than in the nonconducting model B, 
thereby delaying runaway growth until $t\sim24$ Gyr
(for the particular amplitude of initial perturbations). 
The mass inflow rate is correspondingly very small throughout its entire
evolution. Had we started the evolution with a higher initial 
perturbation amplitude,
it would of course take less time to reach a fully nonlinear state.
Small-scale perturbations in real clusters may well have
high amplitudes, but they are readily erased by conduction.
Since it seems unlikely that the amplitudes of perturbations that
are spherically symmetric and coherent over $100$ kpc are more 
than $\sim10\%$ of background quantities,
we believe that galaxy clusters with $f\simgt0.2$ 
are not likely to exhibit a strong thermal instability for many Gyr.

\section{Local Nonradial Modes}

We now carry out a local analysis of nonradial Lagrangian perturbations
in a static, stratified medium. We do not attempt to solve the full 
eigenvalue problem for nonradial modes,
but instead focus on the local behavior of nonradial modes in the presence 
of density and/or temperature stratification. Similar work has been reported 
by \citet{whi87} and \citet{mal87} who used Eulerian perturbations and 
showed that a restoring buoyancy force from a stable entropy gradient can 
change thermal instability into an overstability, confirming the previous 
results of \citet{def70}. Using Lagrangian perturbations, on the other hand, 
\citet{bal88} showed that the negative radial gradient of a net cooling 
function leads to convectively unstable flows in a non-equilibrium system. 
He further claimed that the Eulerian and Lagrangian approaches give different 
results, although in a subsequent paper \citet{bal89} showed that
the Lagrangian approach of thermal instability in dynamical media
produces results that are apparently consistent with 
those from the Eulerian analysis (Balbus 2003, private communication).
We show below that the Lagrangian approach in fact gives exactly the same 
dispersion relation for nonradial thermal instability 
as the Eulerian analyses referred 
to above, provided the background is in initial thermal equilibrium.

We use the Lagrangian technique and 
linearize equations (\ref{eq_cont})-(\ref{eq_flux}) 
assuming that perturbations have small amplitudes of the form
$\sim Y_{lm}(\theta,\phi)e^{ik_r r+\sigma t}$, where 
$Y_{lm}(\theta,\phi)$ is the usual spherical harmonic. 
The detailed procedure to derive a local dispersion relation
is given in the Appendix:
we write only the final result here,
\begin{equation}\label{nonrdisp}
\sigma^3 - \left(\sigma_\infty-
\frac{\gamma-1}{\gamma}\frac{\kappa T}{P}k^2\right)\sigma^2
+ \frac{k_t^2}{k^2}\oBV^2\sigma
+ \frac{\gamma-1}{\gamma P}
\frac{k_t^2}{k^2}
\frac{d\Phi}{dr}
\frac{d}{dr} (\rho\calL + \nabla\cdot\mathbf{F}) = 0,
\end{equation}
where $k_t\equiv [l(l+1)]^{1/2}/r$ is the tangential wavenumber,
$k^2=k_r^2+k_t^2$,  and
\begin{equation}\label{BV}
\oBV^2 \equiv \frac{d\Phi}{dr}\left(\frac{1}{\gamma}\frac{d\ln P}{dr}
-\frac{d\ln\rho}{dr}\right)
\end{equation}
is the \BVf\ frequency for convective motions in a radially stratified 
atmosphere. When the conductivity $\kappa=0$, equation (\ref{nonrdisp}) 
reduces to equation (3.17) of \citet{bal88}\footnote{Since \citet{bal88} 
took $\rho\calL\neq0$ in an initial state, it is unclear how to compute
an effective adiabatic index $\Gamma_1$ in his notation. 
Applying $\Delta$ directly to his equation (2.1c), one obtains 
$\Gamma_1=(5\sigma/2+\tau\Theta_{\tau,P})/
(3\sigma/2 + \Theta + \Theta_{\tau, P})$ 
instead of his equation (2.8). Plugging this expression into his equation 
(3.17) would produce exactly the same local
dispersion relation as our equation (\ref{nonrdisp}) when $\kappa=0$.
Then, the condition $d\Theta/dr<0$ in his equation (3.28) is replaced by
$d(\rho\calL)/dr<0$.}.  If $d(\rho\calL + \nabla\cdot\mathbf{F})/dr=0$
in the last term, equation (\ref{nonrdisp}) becomes exactly the same as
the dispersion relation of the Eulerian approach \citep{mal87}.

It is the last term in equation (\ref{nonrdisp}) that has led to the 
discrepancy between the Eulerian and Lagrangian perturbation analyses. 
This term vanishes if the background is in thermal balance, either in
the presence of thermal conduction in which case 
$\rho\calL + \nabla\cdot\mathbf{F}=0$, or with some other sources of 
heating in which case $\calL = 0$ everywhere. If one perturbs equation 
(\ref{eq_engy}) using the Eulerian operator $\delta$ and Lagrangian 
operator $\Delta$, respectively, and takes the difference of the resulting 
equations, the residual term is 
$\vecxi\cdot\nabla(\rho\calL + \nabla\cdot\mathbf{F})$.
This explains the origin of the last term in equation (\ref{nonrdisp})
and the apparent discrepancy of the Eulerian and Lagrangian approaches.
The two approaches are consistent with each other only if the background 
state is in strict thermal equilibrium. 

One may argue that the assumption of initial thermal equilibrium
is very special and a local analysis can be carried out even
in a non-equilibrium system. However, such an analysis would be misleading 
unless the change of the background state either is allowed for in doing
the perturbations or occurs slowly compared to the growth of perturbations 
of interest. As we showed in \S 3, the cooling time in galaxy clusters is
comparable to the growth time of thermal instability. 
Moreover, in \S4, we performed a numerical simulation of a cluster that
is initially out of thermal equilibrium (model A) and showed that 
the resulting cooling flow quickly dominates the cluster evolution, 
leaving no time for the development of a local thermal instability. 

Since we assume the system is in initial thermal balance, we rewrite 
equation (\ref{nonrdisp}) as
\begin{equation}\label{fdisp}
\sigma^2 - \left(\sigma_\infty-
\frac{\gamma-1}{\gamma}\frac{\kappa T}{P}k^2\right)\sigma
+ \frac{k_t^2}{k^2}\oBV^2 =0,
\end{equation}
which was already derived and discussed by \citet{mal87}. Compared to 
equation (\ref{ldisp}) for radial perturbations, equation (\ref{fdisp}) 
has an extra term that arises from the local buoyancy force. Since 
$dT/dr>0$ and $dP/dr<0$, therefore $\oBV^2>0$, and galaxy clusters are 
stable to convective instability. The buoyancy term has then a stabilizing 
effect, reducing the growth rate of the thermal instability and even 
altering the character of the instability to an overstability when 
$k_t/k$ is high enough \citep{def70,mal87}. If a medium is thermally 
overstable, a bubble that is displaced radially outward (inward) becomes 
more (less) dense compared to the adiabatic case, thus experiencing an
increase of its oscillation amplitude with time.

From equations (\ref{ldisp}) and (\ref{fdisp}), it is trivial to show 
that for a given wave-vector $\mathbf{k}$, Re($\sigma$) of the local 
nonradial mode is always smaller than that of the local radial mode
(e.g., \citealt{whi87}), implying that the latter is more unstable.
This suggests that global nonradial modes, if they exist, would have 
smaller growth rates compared to global radial modes.

\section{Summary and Discussion}

\citet{nar01} and \citet{cho03} have shown that thermal conduction is
quite efficient in a fully turbulent, magnetized plasma.  The
diffusion coefficient in such a medium is approximately a fraction
$f\sim$ a few tenths of the Spitzer coefficient (the full Spitzer coefficient
applies to an unmagnetized medium). Conduction at this level is sufficient 
to provide significant heat to the cooling gas in a galaxy cluster and to 
maintain energy balance in the cluster center 
(\citealt{nar01,voi02,fab02}; ZN03).
This is an attractive explanation for the suppression of mass dropout in
cooling cores of clusters.
However, the demonstration that one can build equilibrium models with
heating equal to cooling is not sufficient, since the hot gas is
likely to be subject to a thermal instability \citep{fie65}.  To be
fully consistent with the new X-ray data on clusters, which show the absence 
of gas below a few keV, it is necessary to demonstrate that the thermal
instability is either fully suppressed or is at least much
weaker than expected.

It is well-known that thermal conduction suppresses thermal
instability in a hot gas for perturbations with short wavelengths.
Indeed, using a local WKB-like analysis, ZN03 have shown that
perturbations with wavelengths up to a fraction of the radius are
thermally stable in clusters.  However, this still leaves open the
possibility that one or more long-wavelength global modes may be
unstable.  In this paper, we have analyzed in some detail the
stability of such global modes.

Following ZN03, we set up the equilibrium density and temperature
profiles of clusters, assuming strict hydrostatic and energy
equilibrium (\S2).  By applying Lagrangian perturbations, we derive a
set of differential equations that describe the eigenvalue problem for
global radial modes, with the mode growth rate $\sigma$ acting as the
eigenvalue (eqs.\ [\ref{eq_f1}]--[\ref{eq_f3}], \S3).  
In the absence of conduction ($f=0$),
we find as expected that a cluster has an infinite number of unstable
modes, all with rapid growth rates $\sigma\sim \sigma_\infty=
1/t_\infty$ (eqs.\ [\ref{ldisp}] and [\ref{eq_tinf}]), 
where $t_\infty$ is the standard growth
time scale associated with the thermal instability.  For typical
cooling flow clusters, the modes should grow in less than a Gyr,
implying a serious threat to stability.  However, when we set $f=0.2$,
roughly the value recommended by \citet{nar01} and ZN03, we find that
all radial modes except one become stable.  The one residual unstable
mode is the fundamental nodeless (in $\xi_r/r$) mode, i.e. the mode with 
the longest possible ``wavelength'' that can fit within the system.  
This mode is weakly unstable, with a growth time scale $t_{\rm grow}$ that 
is 6 to 9 times longer than $t_\infty$. \citet{sok03} has recently posted a
paper in which he describes a quasi-global stability analysis of
clusters.  However, both the initial equilibrium he assumes and the
perturbations he considers are very approximate, and it is hard to
compare his results with ours.

For the clusters we have studied, $t_{\rm grow}$ for the lone unstable
mode is $\sim2-5$ Gyr (Table 1).  This is an interestingly long time
scale, which is probably comparable to the elapsed time since the last
major merger event in a hierarchical clustering scenario (typically
$\sim7$ Gyr for massive clusters, e.g., \citealt{kit96}).  We imagine
that major mergers (and perhaps to a lesser extent even minor mergers)
leave the cluster gas in a highly mixed and turbulent state.  If
the merger drives the whole system well out of thermal equilibrium, 
mass dropout would occur very rapidly ($<1$ Gyr) at the central parts 
(as in model A in \S4), which may in turn allow the remaining
gas to achieve a new thermal equilibrium that is similar to its present state.
In a sense, the merger acts to reset initial conditions.  
After the merger, the unstable global radial mode would grow.
However, if the growth time for the mode is comparable to the
effective age of the cluster since the last merger, as appears to be
the case for our models, then we do not expect the thermal
instability to be a major problem.

We have confirmed the results of \S3 by means of numerical simulations
of radial perturbations in a model cluster (\S4, Fig. 5).  Three
models are discussed.  Model A considers a cluster without any heating
to balance cooling.  This model exhibits the classic cooling flow
catastrophe, a violent and very rapid runaway at the cluster center.
The runaway is so rapid that there is no possibility for the thermal
instability to do anything.  Model C considers a cluster in which
cooling is exactly balanced by conductive heating in equilibrium.  The
numerical simulation shows that this model has a slow instability that
grows at precisely the rate calculated in \S3.  For the particular
initial conditions selected, the mode grows to the non-linear runaway
stage only after 24 Gyr, which means that the model is for all
practical purposes stable.  Model B is an interesting in-between case
in which heating and cooling are balanced, except that the heating is
not from conduction but from some other local (unspecified) agency.
This model shows the classic thermal instability.  The perturbations
grow quite rapidly, on the time scale $t_\infty$ of the \citet{fie65}
instability. In 4 Gyr, the model is completely destroyed by
nonlinear runaway.

We thus reach the following important conclusion from the simulations:
apart from balancing heating and cooling, it is necessary also to make
sure that the heating is of the right kind to control the thermal
instability.  Specifically, the heating must involve diffusive
transport of energy so that short-wavelength perturbations are
smoothed out and not allowed to grow, and even long-wavelength
perturbations are partially stabilized.  Thermal conduction is 
diffusive in nature and is quite effective in this respect,
as we have shown in this paper. Turbulent mixing would be similarly effective.
AGN heating via a jet might behave diffusively if the heat is transferred to 
the cluster gas via mechanical turbulence. Heating through mass infall in 
minor mergers might also be effective since in this case again the heat is 
likely to spread through the gas via turbulence.
However, radiative heating \citep{cio01} or cosmic ray heating \citep{loe91} 
from an AGN are unlikely to control the thermal instability, since these 
heating mechanisms are not diffusive in nature.  

In this connection, we note that not all solutions to the cooling flow
problem involve equilibrium models.  \citet{kai03} have
described an interesting model involving AGN heating in which the gas
goes through a limit cycle.  For most of the time, the AGN at the
center is in quiescence and the gas is not significantly heated.  As
the gas undergoes a cooling catastrophe, the gas density around the
AGN increases, causing the AGN to switch on, to eject a powerful jet
and to heat the cluster gas.  The heated gas expands, the AGN switches
off, and the gas starts the cycle again. The analysis we have carried
out in this paper does not include such scenarios, since we have
explicitly assumed that heating and cooling are balanced.

In \S5, we analyze the properties of nonradial perturbations in a
cluster.  This problem has been studied by \citet{def70},
\citet{whi87}, and \citet{mal87} using an Eulerian approach, and by
\citet{bal88} using a Lagrangian approach.  We show that the
Lagrangian analysis leads to the same local dispersion relation (see
eq.\ [\ref{fdisp}]) as the previous Eulerian analyses, provided that
the background is in thermal equilibrium.  For a cluster that is
convectively stable (as all clusters are), $\oBV^2>0$ (eq.\
[\ref{BV}]), and the buoyancy force associated with the entropy
gradient plays a stabilizing role. As a result, if we consider radial
perturbations with a given wave-vector that are thermally unstable,
then non-radial perturbations with the same wave-vector always have a
lower growth rate.  This means that the most unstable mode is always a
radial mode.  Although this result is obtained via a local analysis,
it is expected to be true for global modes as well.  We have,
therefore, not analyzed global non-radial modes.

Is it an accident that the mode growth time that we have estimated is
comparable to the age of the system as measured since the last major
merger?  We suggest that perhaps it is not.  Imagine a cluster that is
formed with sufficient gas such that several modes are initially thermally
unstable with fairly short growth times.  If the cluster were to start
from a well-stirred initial state (say immediately after a merger), the
various modes would grow and when the age of the system exceeds the growth 
time of any particular mode, that mode would go non-linear and cause a 
certain amount of mass to cool and drop out from the hot medium. 
The mass dropout will cause the 
amount of gas in the hot phase to decrease, and as a result, the other
modes would become less unstable. With continued mass dropout, perhaps
only one unstable mode would be left finally. This mode would remain in the 
system so long as its growth time is not much shorter than the current age. 
With increasing age, presumably this mode too will cause some mass dropout, 
but always in such a manner that the gas that is left will have a growth time 
comparable to the age. In this picture, mass dropout acts as a safety valve 
that enables the system to be always in a state of marginal equilibrium.
If this scenario is valid, then it is of course natural that the clusters we
observe have unstable modes with growth times of order several Gyr.

In this paper, we have assumed the parameter $f$ to be constant over the 
entire volume of a cluster. This may not be a good approximation in some 
clusters that exhibit sharp discontinuities in temperature and density 
(e.g., \citealt{mar00,vik01a,vik01b,dup03}). 
These cold fronts, located at about $r\sim300$ kpc, probably result 
from cluster mergers. \citet{vik01b} argued that strong magnetic fields 
parallel to cold fronts may be responsible for the low conductivity.
If so, the existence of cold fronts is unlikely to affect the strong 
stabilizing role played by conduction for the bulk of the cluster gas. 
The cold fronts are presumably transient features, surviving for only a
dynamical time ($<1$ Gyr) before being 
disrupted by Kelvin-Helmholtz instability \citep{maz02} or by merger shocks 
\citep{nag03}. 

From spatial variations of temperature detected in the 
cluster A754, \citet{mar03} argued that, in addition to cold fronts, 
the bulk of the gas in this cluster may have conductivity much smaller 
than the Spitzer value. The critical question is how recently were the
observed temperature inhomogeneities formed and how much longer will
they last. If the cluster is undergoing a merger, for instance, the density
and temperature fluctuations are all transient. Although conduction tries 
to smooth out local temperature fluctuations, these structures may be 
continuously created by subsequent mergers and/or the sloshing motion
of the gas in the dark matter potential \citep{roe98}.
Acoustic motions of the gas may also induce temporary fluctuations.

Since a large fraction of galaxy clusters exhibit powerful extended
radio sources at their centers, many authors have considered outbursts
from the central AGN to be the source of heat to balance the radiative
cooling in clusters.  As noted above, this works best if the heating
is done via some turbulent agency so that short wavelength thermal
instability can be controlled.  In addition, the AGN heating cannot be too 
strong since it would smooth out and erase the radial variations of 
metallicity that have been observed in some clusters 
(e.g., \citealt{all01,joh02}).

A very real possibility is that both AGN heating and conductive
heating work together in clusters, though their relative importance
may vary from cluster to cluster (ZN03).  It is even possible that AGN
heating dominates in the inner regions, while conduction plays a more
important role farther out.  Recently, \citet{rus02} and \citet{bri03}
demonstrated that simultaneous heating by AGN and thermal conduction
produces quasi-static density and temperature profiles that are
similar to those of observed clusters. In particular, \citet{bri03}
showed that clusters with AGN heating alone exhibit either too high
mass accretion rates or unrealistic temperature distributions, while
conductive heating with $f\sim0.35$ alone or together with AGN heating
gives reasonable fits to observations.  Although the effects of their
computational methods (e.g., removal of cold gas out of the 
computational domain ) and prescriptions for the AGN heating are 
uncertain, their simulations suggest that the stabilization of the hot
gas is most likely achieved by conduction. Therefore, even in clusters 
where AGN heating is dominant, the role of thermal conduction as a 
stabilizing agent should not be ignored (ZN03). 

\acknowledgements We gratefully acknowledge helpful discussions with
S.\ Balbus, L.\ David, G.\ Field, E.\ Ostriker, and N.\ Yoshida. 
We also thank an anonymous referee for useful comments.
This work was supported in part by NASA grant NAG5-10780.

\appendix

\section{Nonradial Modes}
\label{apdx1}

In this appendix, we linearize equations (\ref{eq_cont})-(\ref{eq_flux})
for nonradial modes of perturbations and derive a local dispersion
relation for thermal instability in a stratified medium. 
As in \S3 for radial modes, our approach is Lagrangian and includes the 
effects of thermal conduction. Similar analyses have been performed 
by \citet{mal87} for Eulerian perturbations, and by \citet{bal88}
for Lagrangian perturbations, but without conduction.
The effects of a radial mass flow and magnetic fields were studied 
by \citet{bal89} and \citet{bal91} using Lagrangian perturbations.
We apply $\Delta$ to equations (\ref{eq_mome}) and
(\ref{eq_engy}) and obtain
\begin{equation}\label{app_mome}
\frac{d^2\vecxi}{dt^2} = (\nabla\cdot\vecxi)\nabla\Phi
+ \frac{1}{\rho}\nabla(P\nabla\cdot\vecxi)
- \frac{1}{\rho}\nabla\left(P\frac{\DT}{T}\right) - 
  \nabla(\vecxi\cdot\nabla\Phi),
\end{equation}
\begin{eqnarray}\label{app_engy}
\left( \frac{P}{\gamma-1}\frac{d}{dt} + \rho T \calL_T \right)
\frac{\DT}{T} & + &
\left(P\frac{d}{dt} - \rho^2\calL_\rho\right) (\nabla\cdot\vecxi)
\nonumber \\
& = &\frac{1}{4\pi r^2}\frac{\partial}{\partial r} \DLr
+ \kappa \nabla_t^2 \left(\Delta T - \xi_r \frac{dT}{dr}\right)
- (\nabla_t\cdot\vecxi_t)\nabla\cdot\mathbf{F},
\end{eqnarray}
where $\vecxi_t$ denotes the tangential component of $\vecxi$ and
$\nabla_t$ is the tangential gradient
\begin{equation}
\nabla_t \equiv \frac{\hat{\theta}}{r}\frac{\partial}{\partial\theta}
+ \frac{\hat{\phi}}{r\sin\theta}\frac{\partial}{\partial\phi},
\end{equation}
with $\hat{\theta}$ and $\hat{\phi}$ denoting unit
vectors in the polar and azimuthal directions, respectively.
Equation (\ref{eq_flux_r}) linking $\DT$ and $\DLr$ is valid for 
both radial and nonradial perturbations.

Following \citet{bal88}, we assume that the perturbations are of the form 
(see also \citealt{cox80})
\begin{equation}\label{sph}
\vecxi = [\hat{r}\xi_r(r)+r\xi_t(r)\nabla_t] Y_{lm}(\theta,\phi)e^{\sigma t},
\end{equation}
where $Y_{lm}(\theta,\phi)$ is the spherical harmonic, satisfying 
an identity
\begin{equation}
\nabla_t^2 Y_{lm} = -\frac{l(l+1)}{r^2}Y_{lm}.  
\end{equation}
We assume the same angular and temporal dependences in the other perturbed 
variables $\DT$ and $\DLr$; in what follows, 
we omit $Y_{lm}e^{\sigma t}$ from all the perturbation variables.

The tangential component of equation (\ref{app_mome}) is integrated to give
\begin{equation}\label{tan_xi}
\left[\frac{\rho}{P}\sigma^2 + \frac{l(l+1)}{r^2}\right] r\xi_t
= \frac{1}{r^2}\frac{dr^2\xi_r}{dr} - \frac{\DT}{T}
+\frac{d\ln P}{dr}\xi_r.
\end{equation}
From equations (\ref{sph}) and (\ref{tan_xi}), we thus write
\begin{eqnarray}\label{div_xi}
\nabla\cdot\vecxi & = &
\frac{1}{r^2}\frac{dr^2\xi_r}{dr} - \frac{l(l+1)}{r}\xi_t
\nonumber \\
& = & \frac{1}{D}
\left[\frac{1}{r^2}\frac{dr^2\xi_r}{dr} + 
(D-1) \left(\frac{\DT}{T}-\frac{d\ln P}{dr}\xi_r\right)\right],
\end{eqnarray}
where the dimensionless parameter $D$, defined by
\begin{equation}\label{def_D}
D\equiv 1 + \frac{l(l+1)P}{r^2\rho\sigma^2},
\end{equation}
measures approximately the square of the ratio of the growth time 
to the sound crossing time across a tangential wavelength.

Substituting equations (\ref{sph}) and (\ref{div_xi}) into the radial 
component of equation (\ref{app_mome}), we obtain
\begin{eqnarray}\label{rad_p}
\frac{d^2}{dr^2}\left(\frac{\xi_r}{r}\right)
& + & \left(\frac{4}{r} + \frac{d\ln P/D}{dr}\right)
\frac{d}{dr}\left(\frac{\xi_r}{r}\right) 
\nonumber \\ & + & 
\left[\frac{\rho}{P}
\left(\frac{1}{r}\frac{d\Phi}{dr}-4\pi G\rhodm - D\sigma^2 \right)
+(D-1)\frac{d\ln P}{dr}\frac{d\ln T}{dr}
-\frac{d\ln r^3P}{dr} \frac{d\ln D}{dr} 
\right] \frac{\xi_r}{r}  
\nonumber \\ & = & 
\frac{1}{r} \left[\frac{d}{dr}\left(\frac{\DT}{T}\right)
+ \left(D\frac{d\ln P}{dr} - \frac{d\ln D}{dr}\right)\frac{\DT}{T}
\right].
\end{eqnarray}
Similarly, equation (\ref{app_engy}) becomes
\begin{eqnarray}\label{eng_p}
\frac{1}{4\pi r^2}\frac{d}{dr} \DLr
& = &\left( \frac{P\sigma}{\gamma-1} + \rho T \calL_T \right) \frac{\DT}{T}  
+\frac{l(l+1)}{r^2}\left(\kappa \DT + F_r\xi_r\right) 
-\frac{\nabla\cdot\mathbf{F}}{r^2}\frac{dr^2\xi_r}{dr}
\nonumber \\
& + & D^{-1}(P\sigma - \rho^2\calL_\rho -\rho\calL)
\left[\frac{1}{r^2}\frac{dr^2\xi_r}{dr}
+(D-1)\left(\frac{\DT}{T}-\frac{d\ln P}{dr}\xi_r\right)
\right].
\end{eqnarray}
Note that when $D=1$, corresponding to pure radial modes, 
equations (\ref{rad_p}) and (\ref{eng_p}) are reduced to
equations (\ref{eq_f1}) and (\ref{eq_f2}), respectively.
Subject to proper boundary conditions, equations (\ref{rad_p}), 
(\ref{eng_p}), and (\ref{eq_f3}) may be integrated to yield
solutions for global nonradial modes, but this is beyond the
scope of the present paper.

Let us define the local tangential wavenumber $k_t\equiv [l(l+1)]^{1/2}/r$
and the local radial wavenumber $k_r\equiv d\ln\chi/dr$,
where $\chi$ refers to any perturbed variable. 
Let us also focus on local modes which vary rapidly in both 
radial and tangential directions and grow slowly
compared to the sound crossing time across their wavelengths, i.e.\
$k_rr\gg1$, $k_r(d\ln P/dr)\gg1$, and 
$ k_r^2 \sim k_t^2 \gg \rho\sigma^2/P$ (or $D\gg1$).
Using equation (\ref{eq_f3}) in the following alternative form,
\begin{equation}
\frac{\DLr}{4\pi r^2} = \frac{d}{dr}\left(
\kappa\DT + F_r\xi_r\right) - \xi_r (\nabla\cdot\mathbf{F}),
\end{equation}
we simplify equations (\ref{rad_p}) and (\ref{eng_p}) to
\begin{equation}\label{simp1}
-\left(k^2 - D\frac{d\ln P}{dr}\frac{d\ln T}{dr}\right) \xi_r
= \left(D\frac{d\ln P}{dr}\right)\frac{\DT}{T},
\end{equation}
\begin{equation}\label{simp2}
 - \left[k^2F_r +  \frac{d}{dr}\nabla\cdot\mathbf{F}
 -  (P\sigma-\rho^2\calL_\rho-\rho\calL)
\frac{d\ln P}{dr}\right]\xi_r 
= \left[\frac{\gamma P}{\gamma-1}(\sigma -\sigma_\infty)
+ \kappa Tk^2\right]\frac{\DT}{T},
\end{equation}
where $k^2\equiv k_r^2+k_t^2$ is the amplitude of the 
total wavenumber and $\sigma_\infty$ is defined by equation (\ref{eq_siginf}).
Combining equations (\ref{simp1}) and (\ref{simp2}),
we finally obtain equation (\ref{nonrdisp}) as a local dispersion relation 
for nonradial thermal perturbations.

\clearpage
\begin{deluxetable}{ccccccc}
\tablecaption{Parameters and Time Scales for Five Clusters.
\label{table1}}
\tablewidth{0pt}
\tablehead{
\colhead{Name}                                  &
\colhead{$T(0)$ (keV)\tablenotemark{a}}          &
\colhead{$n_e(0)$ (cm$^{-3}$)\tablenotemark{a}}&
\colhead{$f$\tablenotemark{a}                }  &
\colhead{$\tcool$ (Gyr)\tablenotemark{b}     }  &
\colhead{$\tgrow$ (Gyr)\tablenotemark{c} }      &
\colhead{$\tgrow/t_{\infty,0}$\tablenotemark{c,d}}  
}
\startdata
Abell 1795      &2    &0.049  & 0.2 & 0.98 & 4.1 & 6.3 \\ 
Abell 1835      &5    &0.17   & 0.4 & 0.45 & 1.9 & 6.3 \\ 
Abell 2199      &1.6  &0.074  & 0.4 & 0.58 & 3.5 & 9.0 \\
Abell 2390      &4    &0.069  & 0.3 & 0.98 & 4.6 & 7.1 \\ 
RXJ 1347.5-1145 &6    &0.11   & 0.3 & 0.76 & 3.3 & 6.2 \\ 
\enddata

\tablenotetext{a}{Adopted from ZN03.}
\tablenotetext{b}{$\tcool$ is the isobaric cooling time at the cluster center
(see eq.\ [\ref{eq_tcool}]).}
\tablenotetext{c}{$\tgrow$ is the growth time of the global radial mode.}
\tablenotetext{d}{$t_{\infty,0}$ is the growth time of 
   the local isobaric thermal instability at the cluster center
   in the absence of conduction (see eq.\ [\ref{eq_tinf}]).}
\end{deluxetable}

\clearpage

\clearpage

\clearpage

\clearpage

\clearpage

\clearpage
 

\begin{thebibliography}{}
\bibitem[Afshordi \& Cen(2002)]{afs02}
   Afshordi, N., \& Cen, R.\ 2002 \apj, 564, 669
\bibitem[Allen et al.(2001)]{all01}
   Allen, S.\ W., Ettori, S., \& Fabian, A.\ C.\ 2001, \mnras, 324, 877
\bibitem[Balbus(1986)]{bal86}
   Balbus, S.\ A.\ 1986, \apj, 303, L79
\bibitem[Balbus(1988)]{bal88}
   Balbus, S.\ A.\ 1988, \apj, 328, 395
\bibitem[Balbus(1991)]{bal91}
   Balbus, S.\ A.\ 1991, \apj, 372, 25
\bibitem[Balbus \& Soker(1989)]{bal89}
   Balbus, S.\ A., \& Soker, N.\ 1989, \apj, 341, 611
\bibitem[Bertschinger \& Meiksin(1986)]{ber86}
   Bertschinger, E., \& Meiksin, A.\ 1986, \apj, 306, L1
\bibitem[Binney \& Cowie(1981)]{bin81}
   Binney, J., \& Cowie, L.\ L.\ 1981, \apj, 247, 464
\bibitem[B\"{o}hringer et al.(2001)]{boh01}
   B\"{o}hringer, H., Belsole, E., Kennea, J., Matsushita, K.,
   Molendi, S., Worrall, D.\ M., Mushotzky, R.\ F., Ehle, M.,
   Guainazzi, M., Sakelliou, I., Stewart, G., Vestrand, W.\ T.,
   Dos Santos, S.\ 2001, \aap, 365, L181
\bibitem[Bregman \& David(1988)]{bre88}
   Bregman, J.\ N., \& David, L.\ P.\ 1988, \apj, 326, 639
\bibitem[Brighenti \& Mathews(2002)]{bri02}
   Brighenti, F., \& Mathews, W.\ G.\ 2002, \apj, 573, 542
\bibitem[Brighenti \& Mathews(2003)]{bri03}
   Brighenti, F., \& Mathews, W.\ G.\ 2003, \apj, 587, 580
\bibitem[Br\"uggen \& Kaiser(2002)]{bru02}
   Br\"uggen, M., \& Kaiser, C.\ R.\  2002, Nature, 418, 301
\bibitem[Carilli \& Taylor(2002)]{car02}
   Carilli, C.\ L., Taylor, G.\ B.\ 2002, \araa, 40, 319
\bibitem[Chandran \& Cowley(1998)]{cha98}
   Chandran, B.\ D.\ G., \& Cowley, S.\ C.\ 1998, Phys.\ Rev.\ Lett., 80, 3077
\bibitem[Chandran et al.(1999)]{cha99}
   Chandran, B.\ D.\ G., Cowley, S.\ C., Ivanushkina, M.,
   \& Sydora, R.\ 1999, \apj, 525, 638
\bibitem[Cho et al.(2003)]{cho03}
   Cho, J., Lazarian, A., Honein, A., Knaepen, B., Kassinos, S., \& 
   Moin, S.\ 2003, \apj, 589, L77
\bibitem[Chun \& Rosner(1993)]{chu93}
   Chun, E., \& Rosner, R.\ 1993, \apj, 408, 678
\bibitem[Churazov et al.(2002)]{chu02}
   Churazov, E., Sunyaev, R., Forman, W., \& B\"ohringer, H.\ 2002, 
   \mnras, 332, 729
\bibitem[Ciotti \& Ostriker(2001)]{cio01}
   Ciotto, L., \& Ostriker, J.\ P.\ 2001, \apj, 551, 131
\bibitem[Cox(1980)]{cox80}
   Cox, J.\ P.\ 1980, Theory of Stellar Pulsations (Princeton: 
   Princeton Univ.\ Press)
\bibitem[David et al.(1992)]{dav92}
   David, L.\ P., Hughes, J.\ P., \& Tucker, W.\ H.\ 1992, \apj, 394, 452
\bibitem[David et al.(2001)]{dav01}
   David, L.\ P., Nulsen, P.\ E.\ J., McNamara, B.\ R.\, Forman, W., Jones, C., 
   Ponman, T., Robertson, B., \& Wise, M.\ 2001, \apj, 557, 546
\bibitem[Defouw(1970)]{def70}
   Defouw, R.\ J.\ 1970, \apj, 160, 659
\bibitem[Dupke \& White(2003)]{dup03}
   Dupke, R., \& White, R.\ E., III.\ 2003, \apj, 583, L13
\bibitem[Ettori et al.(2002)]{ett02}
   Ettori, S., Fabian, A.\ C., Allen, S.\ W., \& Johnstone, R.\ M.\  
   2002, \mnras, 331, 635
\bibitem[Fabian(1994)]{fab94}
   Fabian, A.\ C.\ 1994, \araa, 32, 277
\bibitem[Fabian et al.(2001a)]{fab01a}
   Fabian, A.\ C., Mushotzky, R.\ F., Nulsen, P.\ E.\ J., \&
   Peterson, J.\ R.\ 2001a, \mnras, 321, L20
\bibitem[Fabian et al.(2001b)]{fab01b}
   Fabian, A.\ C., Sanders, J.\ S., Ettori, S., Taylor, G.\ B.,
   Allen, S.\ W., Crawford, C.\ S., Iwasawa, K, \& Johnstone, R.\ M.\
   2001b, \mnras, 321, L33
\bibitem[Fabian, Voigt, \& Morris(2002)]{fab02}
   Fabian, A.\ C., Voigt, L.\ M., \& Morris R.\ G.\ 2002, \mnras,
   335, L71
\bibitem[Field(1965)]{fie65}
   Field, G.\ B.\ 1965, \apj, 142, 531
\bibitem[Gaetz(1989)]{gae89}
   Gaetz, T.\ J.\ 1989, \apj, 345, 666
\bibitem[Goldreich \& Sridhar(1995)]{gol95}
   Goldreich, P., \& Sridhar, S.\ 1995, \apj, 438, 763
\bibitem[Goldreich \& Sridhar(1997)]{gol97}
   Goldreich, P., \& Sridhar, S.\ 1997, \apj, 485, 680
\bibitem[Gruzinov(2002)]{gru02}
   Gruzinov, A.\ 2002, astro-ph/0203031
\bibitem[Johnstone et al. (2002)]{joh02}
   Johnstone, R.\ M., Allen, S.\ W., Fabian, A.\ C., \& Sanders, J.\ S.\
   2002, \mnras, 336, 299
\bibitem[Kaiser \& Binney(2003)]{kai03}
   Kaiser, C.\ R., Binney, J.\ 2003, \mnras, 338, 837
\bibitem[Kitayama \& Suto(1996)]{kit96}
   Kitayama, T., \& Suto, Y.\ 1996, \apj, 469, 480
\bibitem[Loewenstein, Zweibel, \& Begelman(1991)]{loe91}
   Loewenstein, M., Zweibel, E.\ G., \& Begelman, M.\ C.\ 1991,
   \apj, 377, 392
\bibitem[Malagoli et al.(1987)]{mal87}
   Malagoli, A., Rosner, R., \& Bodo, G.\ 1987, \apj, 319, 632
\bibitem[Malyshkin \& Kulsrud(2001)]{mal01}
   Malyshkin, L., \& Kulsrud, R.\ 2001, \apj, 549, 402
\bibitem[Maoz et al.(1997)]{mao97}
   Maoz, D., Rix, H.-W., Gal-Yam, A., \& Gould, A.\ 1997, \apj, 486, 75
\bibitem[Markevitch et al.(2000)]{mar00}
   Markevitch, M.\, et al.\ 2000, \apj, 541, 542
\bibitem[Markevitch et al.(2003)]{mar03}
   Markevitch, M.\, et al.\ 2003, \apj, 586, L19 
\bibitem[Matsushita et al.(2002)]{mat02}
   Matsushita, K., Belsole, E., Finoguenov, A., \& B\"{o}hringer, H.\
   2002, \aap, 386, 77
\bibitem[Mazzotta et al.(2002)]{maz02}
   Mazzotta, P., Vikhlinin, A., Fusco-Femiano, R., \& 
   Markevitch, M.\ 2002, \apj, 569, L31
\bibitem[McKee \& Begelman(1990)]{mck90}
   McKee, C.\ F., \& Begelman, M.\ C.\ 1990, \apj, 358, 392
\bibitem[Molendi \& Pizzolato(2001)]{mol01}
   Molendi, S., \& Pizzolato, F.\ 2001, \apj, 560, 194
\bibitem[Nagai \& Kravtsov(2003)]{nag03}
   Nagai, D., \& Kravtsov, A.\ V.\ 2003, \apj, 587, 514
\bibitem[Narayan \& Medvedev(2001)]{nar01}
   Narayan, R., \& Medvedev, M.\ V.\ 2001, \apj, 562, L129
\bibitem[Peterson et al.(2001)]{pet01}
   Peterson, J.\ R., Paerels, F.\ B.\ S., Kaastra, J.\ S., Arnaud, M.,
   Reiprich, T.\ H., Fabian, A.\ C., Mushotzky, R.\ F., \& Jernigan, J.\ G.,
   Sakelliou, I.\ 2001, \aap, 365, L104
\bibitem[Peterson et al.(2003)]{pet03}
   Peterson, J.\ R., Kahn, S.\ M., Paerels, F.\ B.\ S., Kaastra, J.\ S.,
   Tamura, T., Bleeker, J.\ A.\ M., Ferrigo, C., \& Jernigan, J.\ G. 2003,
   \apj, 590, 207
\bibitem[Pistinner \& Shaviv(1996)]{pis96}
   Pistinner, S., \& Shaviv, G.\ 1996, \apj, 459, 1
\bibitem[Press et al.(1992)]{pre92}
   Press, W.\ H., Teukolsky, S.\ A., Vetterling, W.\ T., \& Flannery, B.\ P.\ 
   1992, Numerical Recipes in Fortran (Cambridge: Cambridge Univ.\ Press), 838
\bibitem[Rechester \& Rosenbluth(1978)]{rec78}
   Rechester, A.\ B., \& Rosenbluth, M.\ N.\ 1978, Phys.\ Rev.\ Lett., 40, 38
\bibitem[Reynolds, Heinz, \& Begelman(2002)]{rey02}
   Reynolds, C.\ S., Heinz, S., \& Begelman, M.\ C.\ 2002, \mnras, 332, 271
\bibitem[Roettiger et al.(1998)]{roe98}
   Roettiger, K., Stone, J.\ M., \& Mushotzky, R.\ F.\ 1998, \apj, 493, 62
\bibitem[Rosati et al.(2002)]{ros02}
   Rosati, P., Borgani, S., \& Norman, C.\ 2002, \araa, 40, 539
\bibitem[Rosner \& Tucker(1989)]{ros89}
   Rosner, R., \& Tucker, W.\ H.\ 1989, \apj, 338, 761
\bibitem[Ruszkowski \& Begelman(2002)]{rus02} 
   Ruszkowski, M., \& Begelman, M.\ C.\ 2002, \apj, 581, 223
\bibitem[Rybicki \& Lightman(1979)]{ryb79} 
   Rybicki, G.\ B., \& Lightman, A.\ P. 1979, 
   Radiative processes in astrophysics (New York: Wiley)
\bibitem[Sarazin(1988)]{sar88}
   Sarazin, C.\ L.\ 1988, X-ray emissions from clusters of galaxies 
   (Cambridge: Cambridge University Press)
\bibitem[Shapiro \& Teukolsky(1983)]{sha83}
   Shapiro, S.\ L., \& Teukolsky, S.\ A.\ 1983,
   Black Holes, White Dwarfs, and Neutron Stars (New York:
   Wiley-Interscience), pp 127-147
\bibitem[Soker(2003)]{sok03}
   Soker, N.\ 2003, \mnras, 342, 463
\bibitem[Spitzer(1962)]{spi62} 
   Spitzer, L.\ 1962, Physics of Fully Ionized Gases (New York: Interscience)
\bibitem[Stone \& Norman(1992)]{sto92}
   Stone, J.\ M, \& Norman, M.\ L.\ 1992, \apjs, 80, 753
\bibitem[Tamura et al. (2001)]{tam01}
   Tamura, T., Kaastra, J.\ S., Peterson, J.\ R., Paerels, F.\ B.\ S.,
   Mittaz, J.\ P.\ D., Trudolyubov, S.\ P., Stewart, G., Fabian, A.\ C.,
   Mushotzky, R.\ F., Lumb, D.\ H., \& Ikebe, Y.\ 2001, \aap, 365, L87
\bibitem[Tucker \& Rosner(1983)]{tuc83}
   Tucker, W.\ H., \& Rosner, R.\ 1983, \apj, 267, 547
\bibitem[White \& Sarazin(1987)]{whi87}
   White, R.\ E., III, \& Sarazin, C.\ L.\ 1987, \apj, 318, 612
\bibitem[Vikhlinin et al.(2001a)]{vik01a}
   Vikhlinin, A., Markevitch, M., \& Murray, S.\ S.\ 2001a, \apj, 549, L47
\bibitem[Vikhlinin et al.(2001b)]{vik01b}
   Vikhlinin, A., Markevitch, M., \& Murray, S.\ S.\ 2001b, \apj, 551, 160
\bibitem[Voigt et al.(2002)]{voi02}
   Voigt, L.\ M., Schmidt, R.\ W., Fabian, A.\ C., Allen, S.\ W.,
   Johnstone, R.\ M.\ 2002, \mnras, 335, L7
\bibitem[Zakamska \& Narayan(2003)]{zak03}
   Zakamska, N.\ L., \& Narayan, R.\ 2003, \apj, 582, 162 (ZN03)



\end{thebibliography}
\end{document}